\documentclass[10pt,journal,compsoc]{IEEEtran}
\ifCLASSOPTIONcompsoc
  \usepackage[nocompress]{cite}
\else
  \usepackage{cite}
\fi
\ifCLASSINFOpdf
\else
\fi
\usepackage{color}
\usepackage{soul}
\usepackage{multirow}
\usepackage{rotating}
\usepackage{amssymb}
\usepackage{amsmath}
\usepackage{algorithmicx}
\usepackage{wrapfig}
\usepackage[normalem]{ulem}

\definecolor{turquoise}{cmyk}{0.65,0,0.1,0.1}
\definecolor{purple}{rgb}{0.65,0,0.65}
\definecolor{darkgreen}{rgb}{0.0, 0.5, 0.0}
\definecolor{darkred}{rgb}{0.5, 0.0, 0.0}
\definecolor{darkblue}{rgb}{0.0, 0.0, 0.5}
\definecolor{blue}{rgb}{0.0, 0.0, 1.0}

\newcommand{\erase}[1]{}

\newcommand{\hide}[1]{{}}

\newcommand{\eg}{{\textit{e.g., }}}
\newcommand{\ie}{{\textit{i.e., }}}

\definecolor{darkblue}{rgb}{0.0, 0.2, 0.5}

\begin{document}
%
\title{PTRM: Perceived Terrain Realism Metrics}
%
%
%
%

\author{Suren Deepak Rajasekaran,
        Hao Kang,
        Bedrich~Benes~\IEEEmembership{IEEE Senior Member},\\
        Martin \v{C}ad\'{\i}k,
        Eric Galin,
        Eric Gu\'erin,
        Adrien Peytavie,
        and Pavel Slav\'{\i}k
\IEEEcompsocitemizethanks{
\IEEEcompsocthanksitem S.~D. Rajasekaran, H.~Kang and B.~Benes were with Purdue Univ, USA
\IEEEcompsocthanksitem M. \v{C}ad\'{\i}k is with FIT, Brno Univ. of Technology and FEL, Czech Technical University, Czech Republic
\IEEEcompsocthanksitem E. Galin, E. Gu\'erin, and A. Peytavie were with Universit\'e de Lyon, France
\IEEEcompsocthanksitem P. Pavel Slav\'{\i}k was with FEL, Czech Technical Univ., Czech Republic
}
\thanks{Manuscript received September 9, 2019.}
}

%
%

{}

%



\IEEEtitleabstractindextext{
\begin{abstract}
Terrains are visually important and commonly used in computer graphics. While many algorithms for their generation exist, it is difficult to assess the realism of a generated terrain. This paper presents a first step in the direction of perceptual evaluation of terrain models. We gathered and categorized several classes of real terrains and we generated synthetic terrains by using methods from computer graphics. We then conducted two large studies ranking the terrains perceptually and showing that the synthetic terrains are perceived as lacking realism as compared to the real ones. Then we provide insight into the features that affect the perceived realism by a quantitative evaluation based on localized geomorphology-based landform features (geomorphons) that categorize terrain structures such as valleys, ridges, hollows, etc. We show that the presence or absence of certain features have a significant perceptual effect. We then introduce \textit{Perceived Terrain Realism Metrics} (PTRM); a perceptual metrics that estimates perceived realism of a terrain represented as a digital elevation map by relating distribution of terrain features with their perceived realism. We validated PTRM on real and synthetic data and compared it to the perceptual studies. To confirm the importance of the presence of these features, we used a generative deep neural network to transfer them between real terrains and synthetic ones and we performed another perceptual experiment that further confirmed their importance for perceived realism.
\end{abstract}

\begin{IEEEkeywords}
I.3. Computer Graphics; I.3.7 Three-Dimensional Graphics and Realism; O.2.3 Perceptual models
\end{IEEEkeywords}
}


\maketitle

\IEEEdisplaynontitleabstractindextext

%
\IEEEpeerreviewmaketitle



\begin{figure*}[hbt]
  \centering
   \includegraphics[width=\linewidth]{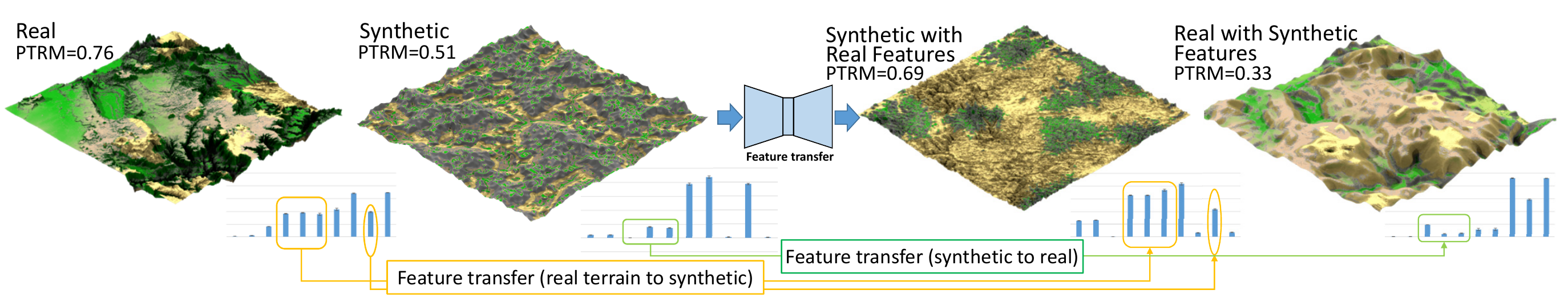}
  \caption{The real terrain from the state of Arizona USA with complex geomorphological patterns has estimated PTRM=0.76 (1=perfect, 0=poor). It has been also ranked by a perceptual as top 78\%. The synthetic terrain models with patterns generated by thermal erosion has PTRM=0.51 and it ranked as 49\% in the study. The corresponding geomorphons show the distribution of patterns in each model with strong presence of valleys, ridges, and hollows landform in real terrain (circled in the graph) that were not so present in the synthetic variety. By using a CycleGAN, we transferred the visually important features to the low-ranked synthetic terrain (orange arrows) and we transferred the features in synthetic terrain to the high-ranked real terrain (green arrow) assuming the real terrain should worsen and the synthetic should improve. The second perceptual study showed that the transferred features improved to PTRM=69 (77\% ranking in our study) and transferring the visually unattractive features from synthetic terrain to the real one demoted its PTRM=0.33 (29\%). The transferred features are circled in the corresponding graphs of geomorphons.}\label{fig:teaser}
\end{figure*}

\section{Introduction}
Terrains are among the most visually stunning structures and their modeling has attracted attention of computer graphics researchers for decades. Patterns in terrains result from eons of complex interacting geomorphological processes with varying strength at differing spatial and temporal scales, which makes them difficult to simulate. 

Humans experience terrains through their entire life and our visual perception system has evolved into a precise tool for judging their realism. Humans are excellent in detecting anomalies~\cite{travers1984human} such as inconsistent rivers, non-realistic shapes of mountains, or incorrectly positioned terrain features, which makes synthetic terrain modeling challenging as quantifying those inconsistencies remains highly complex. Although a wide variety of algorithms exists for modeling terrains (see the recent review of Galin et al.~\cite{Galin19CGF}), existing methods often consider the geomorphological phenomena in separation and their mutual dependencies are neither well-studied nor understood.

Previous methods focused on replicating phenomenological processes of terrain formation, but none, to the best of our knowledge, have focused on the perceived realism of terrain models. The evaluation of results of algorithms simulating natural phenomena has been always a difficult question and is usually addressed by providing side-by-side comparison of the generated structures or is assumed to be correct if the underlying simulations are physically-based.

This paper is a first step in the direction of perceptual validation of realism of computer graphics terrain models. In particular, we attempt to answer the questions: What are the visually important features in terrains that make them realistic? and What is the level of perceived realism of synthetic terrains generated by techniques used in computer graphics? A recent work in geology allows for a quantitative evaluation of terrains by using so called geomorphons that are geomorphological features (valleys, ridges, slopes, spurs, hollows, etc.) that are present in terrain. A~geomorphon is a histograms of features present in a digital elevation map~\cite{JASIEWICZ2013147}. We performed an extensive user study measuring the perceived realism of real and synthetic terrains and we related the realism to geomorphons. We introduce PTRM (Perceived Terrain Realism Metrics) that assigns a normalized value of perceived perception to a terrain represented as a digital elevation model based on the present geomorphons. We validate the PTRM on both real and synthetic terrain models.

Our hypothesis is that some features are visually more important for perceived realism.
We used the state of the art deep neural networks CycleGAN~\cite{zhu:2017:ICCV} to transfer features (valleys, ridges, etc.) from the DEMs that were ranked high to those ranked low and vice versa. We performed another user study that shows that the landforms transferred from highly ranked sets to lowly ranked ones improve the visual perception and that the landforms transferred from low-ranked images to high ranked ones demote them perceptually. Results of the two user-studies combined with the analysis of features show that synthetic terrains do not often include geomorphological features such as \emph{depressions}, \emph{summit}, \emph{flat}, \emph{valley}, \emph{ridge}, \emph{hollow} and \emph{spur}.

An example in Fig.~\ref{fig:teaser} shows a procedural terrain and the distribution of its landform features based on geomorphons as well as a real terrain with its accompanying features. The feature vector of the geomorphons is sorted so that the ones contributing to perceived quality are on the right hand side. The real terrain was ranked as highly realistic (77\%) in our user study and the procedural terrain was on the opposite scale (51\%) as can be also seen in the distribution of the geomorphons. We then used deep learning to transfer the features from the procedural terrain to real and vice versa and we show the corresponding distribution of the geographic features that indicates that the distributions of the geomorphons changed so that the high-ranked worsen and low-ranked improved. This quantitative validation has been then confirmed by a perceptual study that showed that the procedural terrain after the style transfer improves its perceived quality to 69\% and the real terrain worsens to~29\%. We also show the PTRM that predicts how a person would perceive it as realistic (1=perfect, 0=poor).

We claim the following contributions:
1)~we introduce Perceived Terrain Realism Metrics that assigns a normalized value of perceived realism to a terrain represented as a digital elevation model, 2)~we have conducted user studies that validate and measure the perceived realism of real and synthetic terrain models, 3)~we have determined geological features that have effect on perceived reality of terrains, and 4)~we provide a publicly available data-set of real and procedural terrains with assigned perceptual evaluation and calculated geomorphons.

\section{Related Work}\label{sec:rw}
\textit{Perception-based computer graphics approaches:} The knowledge of human perception has been applied in computer graphics since the beginnings and a common way is to incorporate it as a computational model of a particular human visual system (HVS) feature, \eg visual masking~\cite{Ferwerda:1997:MVM:258734.258818}, visual attention and saliency~\cite{7298603, 6751253}, or to fully replace it by a hardware such as an eye tracker~\cite{egst.20041029}. 

Photorealistic rendering traditionally exploits perception limitations to accelerate costly light transport computations~\cite{Weier:2017:PAR:3128975.3129028} and in 3D graphics, HVS models allow removing nonperceptible components~\cite{DBLP:phd/ethos/Reddy97, Reddy:2001:POG:616072.618852} and/or predicting popping artifacts~\cite{Schwarz:2009:PVP:1620993.1621012}. 

Perceptual models have been further applied to improving virtual simulations~\cite{Ondrej:2016:FDA:2974016.2948066}, character animations~\cite{Reitsma:2003:PMC:882262.882304,egst.20041029}, human body modeling~\cite{7935623}, fluid simulations~\cite{um2017perceptual,Bojrab:2013:PIL:2422105.2422107}, and crowd simulations~\cite{Wang:2016:PPA:2856400.2856410,7797248}. High dynamic range imaging and tone mapping benefits from models of human light adaptation~\cite{Mantiuk:2006:PFC:1166087.1166095,Ferwerda:1996:MVA:237170.237262}, color to grey conversions simulate human color sensitivity~\cite{cadik07color_to_gray,smith:inria-00255958}. Interestingness~\cite{6751313} and aesthetic properties of photographs~\cite{6819054}, paintings and fractals~\cite{taylor11,DBLP:journals/cg/SpeharCNT03} have also been approximated by computational models of HVS. 

Close to our work is research on procedural textures~\cite{10.1371/journal.pone.0130335} that aims to define perceptual scales which can steer texture model. The perceived quality of a geometry replaced with texture has also been studied~\cite{Holly2000}. 

Image quality metrics (IQM) utilize HVS models to predict perceptual image quality. 
Full-reference IQMs compute perceptual differences between the reference and distorted images~\cite{Mantiuk:2011:HCV:2010324.1964935, Wang:2004:IQA:2319031.2320551, Wolski:2018:DMP:3278329.3196493}, while no-reference metrics~\cite{NoRM, 6909936} predict the quality in a reference-less setup. Video quality metrics~\cite{4550731, aydin10drivqm} simulate temporal HVS properties to faithfully comparing video sequences. 

Recent research works study perceptual quality of 3D models~\cite{DBLP:journals/tvcg/LavoueLV16} and meshes~\cite{nader16, Guo:2015:ELV:2804408.2804418} including textured models~\cite{Guo:2016:SOV:2997647.2996296}. Visual saliency predictors for 3D meshes have been also proposed~\cite{Wu:2013:MSG:2506575.2506831}.

Unfortunately, no existing metrics is applicable to comparison of synthetic and real terrain images or models, because the compared contents differ significantly. 

\textit{Perception of terrains:} Synthetic terrains have not been studied in perception experiments and we are not aware of any computational perception quality metrics that could be applied. Furthermore, a data-set of synthetic  and real terrains comprising human judgments which could be used for an evaluation of terrain generating methods or for training of data-driven techniques is missing as well. 

Nevertheless, a few research works on classification and perception of real-world terrains have been presented in the fields of environmental psychology and geomorphology.
Dragut and Blaschke~\cite{dragut06} proposed a system for landforms classification on the basis of profile curvature. Several data layers are extracted from the digital terrain model to feed an image segmentation which classifies the terrain into classes like toe slopes, peaks, shoulders, etc. 
Fractal characteristics of terrains were studied in~\cite{HAGERHALL2004247} and they conclude that there is a relationship between preference and the fractal dimension, meaning that fractal dimension may be part of the basis for preference. Finally, scenic beauty and aesthetics have been addressed by~\cite{palmer03,tveit12,Tremblet16,DANIEL2001267}. These works lay the foundation of landscape perception, but they cannot be directly applied to quality assessment of synthetic terrains. Automated tools of measurement and analysis of terrains are sought~\cite{palmer03} to advance this area of research. 

\textit{Terrains in computer graphics} 
have been studied for decades (see the recent review~\cite{Galin19CGF}). Here we list the three major categories of terrain generation techniques: procedural approaches, erosion simulation and by-example.

Historically, the first methods to synthesize terrains relied on procedural and fractal approaches. It consists in finding a way to generate a fractal surface that exhibits self-similarity either by using subdivisions~\cite{Fournier1982,Miller1986}, 
faulting~\cite{Mandelbrot1988}, or by summing noises~\cite{Musgrave1989}. Approaches that control~\cite{Kelley1988} or more specific curve-based constructions~\cite{Gain2009} have been introduced. The overall realism of the generated landscape depends on the fine tuning of control parameters and requires a deep knowledge and understanding of the underlying generation process which restrict those methods to skilled technical artists.

Erosion simulations generate terrain features by approximating
the natural phenomena, such as hydraulic~\cite{Stava:2008,Benes:06:CAVW,Kristof:09} or thermal erosion~\cite{Musgrave1989,Benes:02:WSCG} processes at different scales. 
They are computationally intensive,
and only capture a limited set of small scale structures features~\cite{Cordonier18ToG}, such as ravines or downstream sediment accretion regions. When combined at a larger scale with uplift~\cite{cordonnier2016large}, erosion simulations generate realistic mountain ranges with dendritic ridge networks and their dual drainage network forming rivers. 

Another option to obtain realism by synthesizing
new terrains by-example, for example by stitching together terrain patches  
from existing data-sets. By using techniques from texture synthesis~\cite{Zhou2007} using sparse modeling Gu\'rin et al.~\cite{Guerin2016} generate large terrain with realistic small-scale features. The large scale plausibility remains an open challenge as existing methods, even deep learning~\cite{Guerin2017} oriented approaches, rely on user-sketching and authoring.

Despite recent advances in simulation, the user-control remains an open problem and terrain generation methods only generate a limited set of landforms. Moreover, validation of the generated structures remains an outstanding problem and has been addressed only partially.

\section{Geomorphons}\label{sec:geomorphons}
\begin{figure}[hbt]
  \centering
  \includegraphics[width=\linewidth]{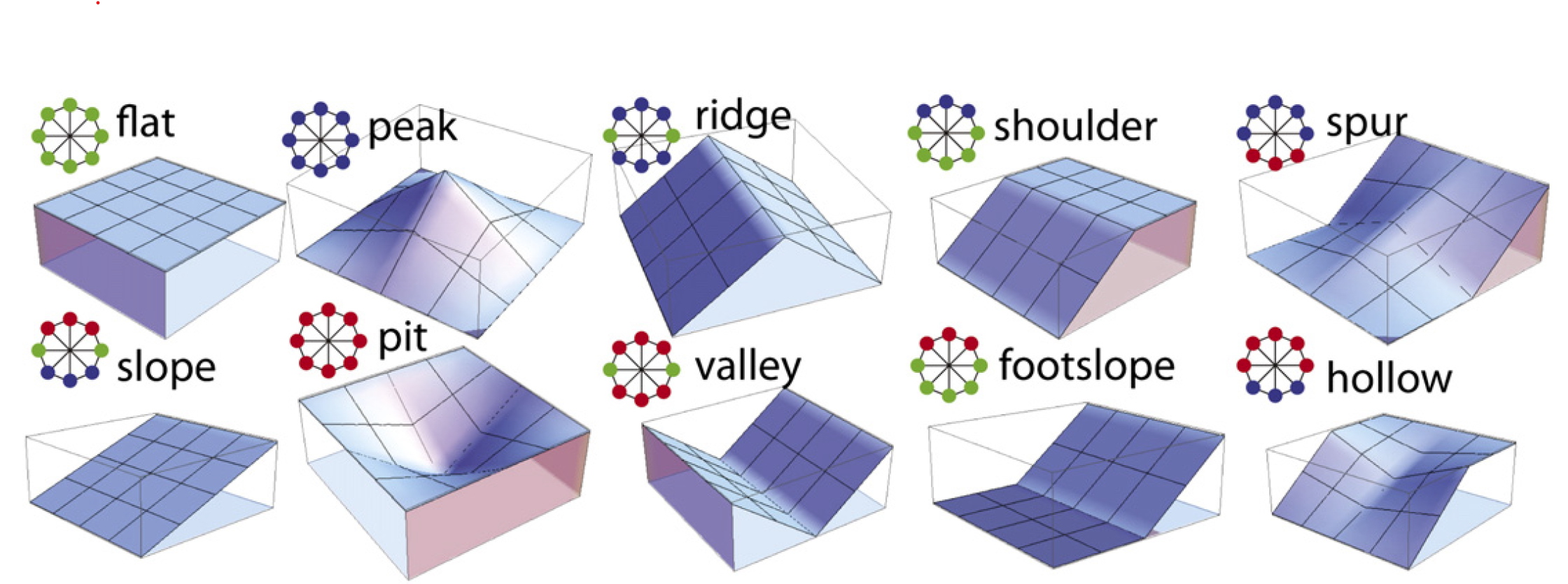}
  \caption{Ten most common land form patterns can be uniquely classified by geomorphons from a DEM. Blue disc identify lower, red higher, and green the same altitude (image from~\cite{JASIEWICZ2013147}).}
  \label{fig:geomorphons}
\end{figure}

The fundamental theory behind our method is the recently introduced concept of \textit{geomorphons}~\cite{JASIEWICZ2013147} that provide an exhaustive classification of terrain features from digital elevation models (DEMs). Geomorphons decompose a DEM into local ternary patterns~\cite{Liao2010} based on the local curvature that provide an oriented eight directional feature vector for each location of the DEM; one value for the Moore neighborhood (see the circles in Fig.~\ref{fig:geomorphons}). This gives rise to ten geomorphons: flat, peak, ridge, shoulder, spur, slope, depression (or pit), valley, footlsope, and hollow, as shown in Fig.~\ref{fig:geomorphons} from~\cite{JASIEWICZ2013147}. Geomorphons depend on the resolution of the DEM, in our setting one pixel of the DEM corresponds to approximately 200m, so each geomorphon describes an area of about $800\times{800}~$m$^2$.

\begin{figure}[hbt]
  \centering
  \includegraphics[width=\linewidth]{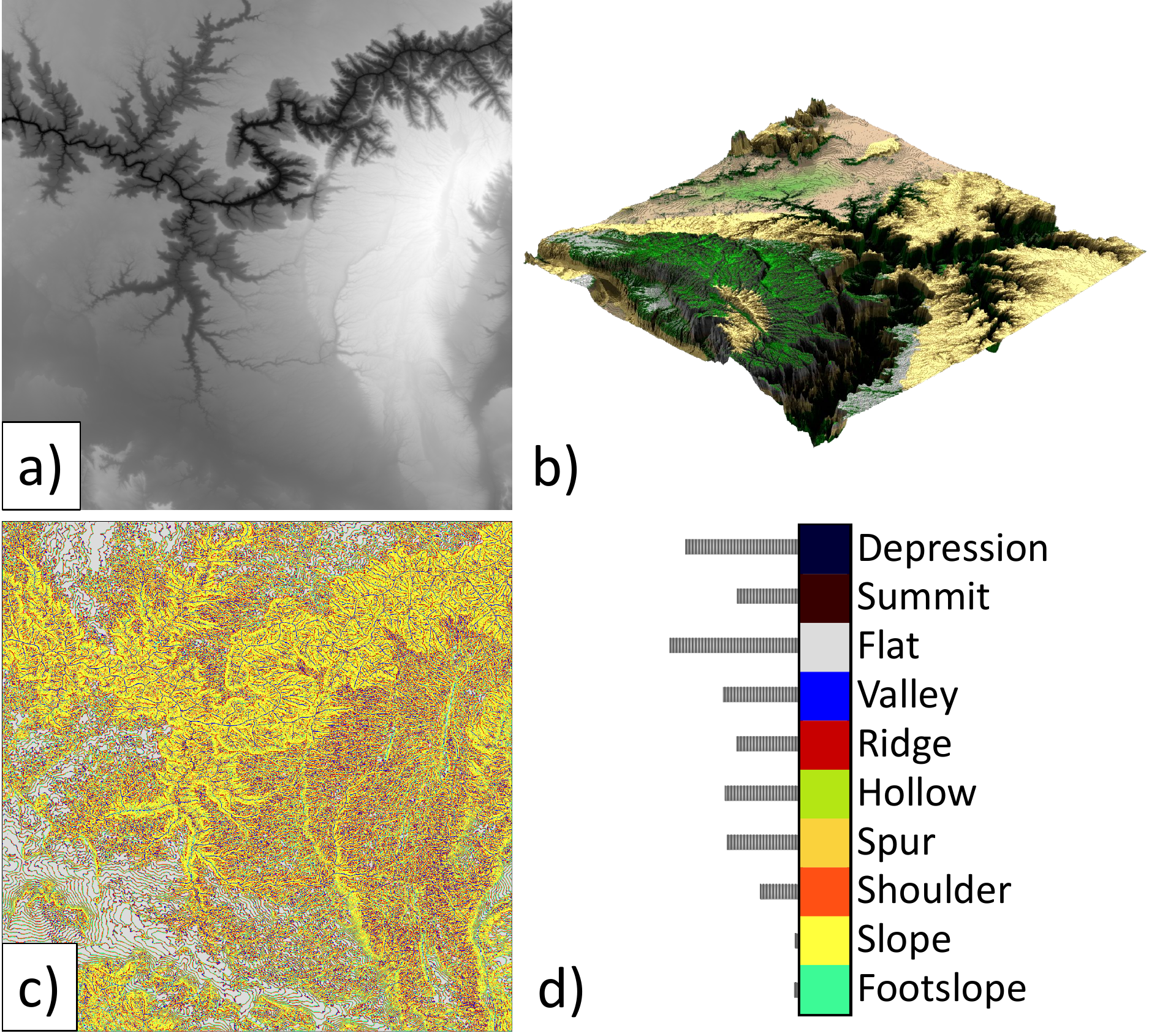}
  \caption{a) The input DEM b) its rendering and c) the geomorphons d) with the explanation of the color-coding.}
  \label{geom-ex}
\end{figure}

We utilize geomorphons to provide understanding of the importance of individual geomorphological landform features and how they affect the perception of terrains. Later we show how they are present or missing in different terrains. The order of the geomorphons in the color coding in Fig.~\ref{geom-ex} is arbitrary and to compare the wide variety of terrains used in this paper, we decided to sort the geomorphons according to their presence in the most realistically perceived terrain category from our user study that are glacial patterns of real terrains (Sec.~\ref{sec:test1}). Fig.~\ref{fig:geomorphons-graph} shows the normalized frequency of geomorphons in all datasets used in this paper and we use the ascending order of geomorphons as: \textit{Depression (or pit)} (the least present), \textit{Summit, Flat, Valley, Ridge, Hollow, Spur, Shoulder, Slope, and Footslope} (the most frequently present).

We used an open implementation of geomorphons in GRASS GIS tool~\cite{neteler2013open} that generates color-coded image corresponding to the input DEM as shown in example in Fig.~\ref{geom-ex}. The output of the algorithm is the normalized coverage of each geomorphon in the input DEM (the values of geomorphons for all datasets from this paper are in the supplemental material).

\begin{figure*}[hbt]
\centering
\includegraphics[width=0.9\linewidth]{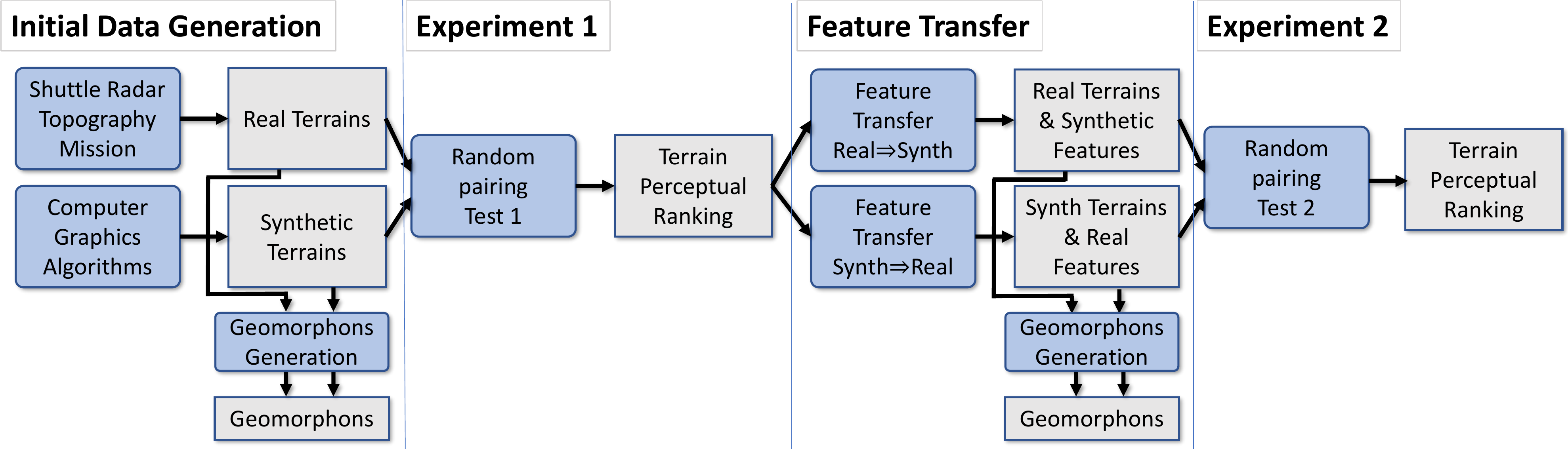}
  \caption{Overview (rounded boxes - processes, squared boxes - data): The \textit{initial data for Experiment 1} were acquired from two sources: real and synthetic, they were rendered and we also generated geomorphons for each image that quantitatively describe their landform features. During the \textit{Experiment  1} we acquired perceptual ranking of each image. The \textit{feature transfer} transferred features from highly ranked images (Real$\Rightarrow$Synth) and vice versa (Synth$\Rightarrow$Real) resulting in two new datasets. The \textit{Experiment 2} perceptually evaluated the initial data and the newly generated ones, confirming that the transferred features have importance on the perceived realism.}
  \label{fig:overview}
\end{figure*}

\section{Method Overview}\label{sec:overview}
The key question we are trying to answer is the perceived realism of terrains and the visual perception and evaluation of synthetic terrains generated by terrain modeling methods in computer graphics. We focus  on the terrain geometry only and we do not consider any additional features such as snow, vegetation, or water bodies. Our work builds on the recent advances in geomorphology, in particular we use the concept of \textit{geomorphons} that are features extracted from Digital Elevation Models (DEMs) that quantitatively measure presence of various shapes in terrain (Sec.~\ref{sec:data}).

We performed two large scale user study (Sec.~\ref{sec:tests}). The first quantifies the perception of real and synthetic data-set and the second one quantifies the effect of the transferred features. Fig.~\ref{fig:overview} shows the overview of our testing. 

During the \textit{initial data generation}, we acquired data of real terrains from Shuttle Radar Topography Mission and we carefully selected several classes featuring prevalent geological patterns (see Tab.~\ref{tab:terrain-numbers}): Aeolian, Coastal, Fluvial, Glacial, and Slope. Then we generated synthetic data-sets by using terrain generation algorithms used in computer graphics: coastal, thermal and fluvial erosion, fractional Brownian motion, noise and ridged-noise terrain models. Geomorphons were generated for each image.

\textit{Experiment 1} (E1) was two-alternative forced choice design -- 2AFC by using Mechanical Turk. We have shown pairs of images and we asked the viewers the question: \textit{"Which terrain looks more realistic (left or right)?"}. Each image received multiple rankings and the number of votes determined its positioning in the overall test. The experiment provided initial terrain ranking for each image and for each image category within each group (real or synthetic). The results were used to construct the PTRM that relates the presence of geomorphons to the perceived realism. In order to evaluate a new terrain, geomorphons need to be calculated, normalized, and input into the PTRM.

\textit{Feature Transfer:} we used the CycleGAN~\cite{zhu:2017:ICCV} to transfer features from the images that were ranked high to those ranked low and conversely (Sec.~\ref{sec:gan}). The motivation for this step is the underlying assumption that certain features have important effect on the visual perception of terrains. This step generated a new data-set that we call S2R (synthetic to real) and R2S (real to synthetic). S2R indicates that procedural features were transferred to the real terrains and R2S is the opposite process. Geomorphons were also generated for the new data-sets.

\textit{Experiment 2} (E2) was also 2AFC, but it included the newly generated sets S2R and R2S (Sec.~\ref{sec:test2}). The underlying assumption was that the features from the highly ranked terrains will be transferred to the low ranked terrains and the new terrains will improve their perceived realism. Similar expectation was hold for the low-ranked terrains, assuming the transferred features would improve their rank. Moreover, for each terrain we also generated the corresponding geomorphons and we kept a careful track of which features were transferred (Sec.~\ref{sec:PTRM}).

\begin{figure*}[h!bt]
\raggedright
\includegraphics[width=\linewidth]{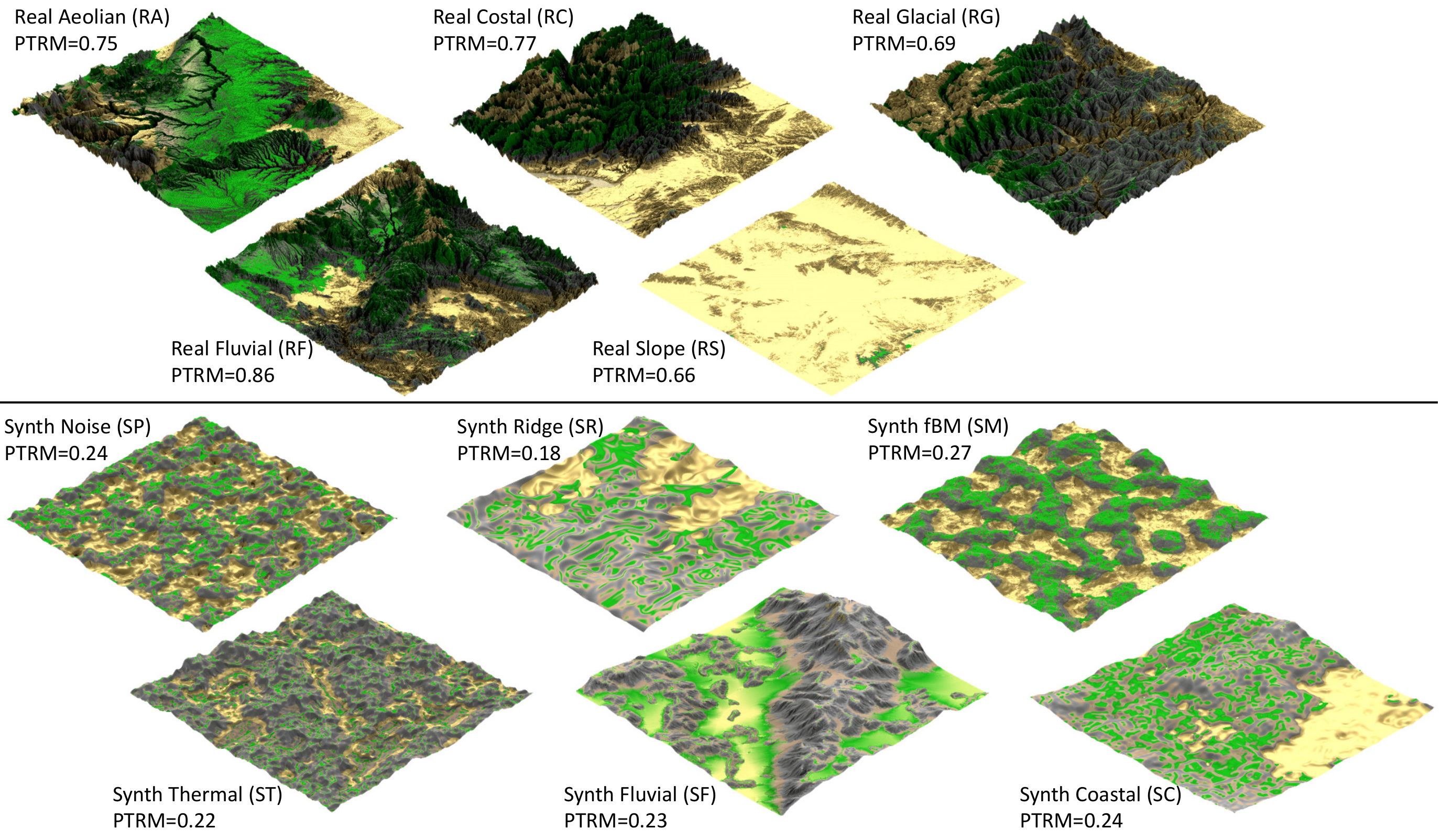}
  \caption{
  (Top) Examples of real terrains rendering used in our experiment and their PTRM: RA)~aeolian patterns from Moab Arches National Park Utah USA, RC)~coastal patterns from Gobi desert Mongolia, RG)~glacial erosion patterns from Himachal Pradesh Western Himalaya India, RF)~fluvial pattern from Chichiltepec Mexico Guerrero, and RS)~slope pattern from Death Valley California USA. (Bottom) Examples of synthetic terrains SP)~noise-based, SR)~ridged-noise, SM)~fractional Brownian motion surface, ST)~thermal erosion, SF)~fluvial erosion, and SC)~coastal erosion (see supplementary material for high-resolution images).}
  \label{fig:synth-real}
\end{figure*}

\section{Terrain Data}\label{sec:data}
The objective of this study was to compare the terrains with the most prevalent geomorphological processes with visually distinguishable features and the common terrain synthesis methods in computer graphics. The DEMs used in this study come from Shuttle Radar Topography Mission (SRTM) data-set~\cite{farr2000shuttle}. We used the three arc-second capture resolution (90m pixel resolution along the equator) tiles from the data set as some of the tiles from the one arc-second (30m pixel resolution along the equator) data that covers the whole globe is not made available to public yet. The resolution roughly translates to $1^{\circ}$ Longitude $\times 1^{\circ}$ Latitude or $100\times{100}$ km resolution approximately depending on the DEM's location on Earth. All the DEMs maintained a resolution of $512\times{512}$ that gives sizes of the land features around $200$ meters per pixel.

\subsection{Real terrains} \label{sec:terrain_data_real}
We used terrains that include patterns that commonly results from aeolian, glacial, coastal, fluvial, and slope processes~\cite{huggett2016fundamentals} along with the retrievability of suggested patterns from the SRTM data-set~\cite{farr2000shuttle}. It is important to note that the geoforming processes are not well-understood and most of the terrains are affected by several of them either at the same time period or in an indeterminable unknown sequence. So instead of discussing processes, we consider terrains that include the specified geomorphological patterns and structures. The two top rows of Fig.~\ref{fig:synth-real} show examples of several renderings of real terrains and the supplementary materials include all data.

\subsection{Synthetic Terrains}\label{sec:terrain_data_synthetic}
We used terrains generated by noise~\cite{Perlin:85:SIGG}, ridged-noise~\cite{Galin19CGF}, fractional Brownian motion (fBm) surfaces~\cite{Fournier1982}, thermal erosion~\cite{Musgrave1989}, fluvial erosion (we used the implementation of ~\v{S}\v{t}ava et al.~\cite{Stava:2008}, but~\cite{Kristof:09,Anh:2007,Neidhold:2005} could be used), and coastal erosion approximated by hydraulic erosion applied only to coastal areas.
Eroded terrains were generated from noise-based terrains (Fig.~\ref{fig:synth-real} two bottom rows).

While procedural generation of terrains is simple so we could have generated an arbitrary number of DEMs, it is rather difficult to find good samples for all the above-mentioned examples of real patterns. Tab.~\ref{tab:terrain-numbers} shows how many terrain models we had for each category and also establishes nomenclature for each set. Each real image starts with the letter R and synthetic with S, the second letter indicates subcategory. We refer to all images from real datasets, R and all synthetic as S. The size of each data-set was the same: $|S|=|R|=150$.
\begin{table}[hbt]
\begin{center}
\begin{tabular}{ |l | l | c| c |}
\hline
\textbf{Type} & \textbf{Category} & \textbf{Abbr.} & \textbf{Sampl.}\\\hline
Real (R)    & Aeolian  & RA & 55\\
            & Coastal  & RC & 19\\
            & Fluvial & RF  & 64\\
            & Glacial & RG  & 07\\
            & Slope   & RS  & 05\\ \hline
Synthetic (S)  & Coastal  & SC & 25 \\
            & fBm     & SM  & 25 \\
            & Fluvial  & SF & 25 \\
            & Noise  & SP  & 25 \\
            & Ridged-noise & SR & 25 \\
            & Thermal  & ST & 25 \\ \hline
Transferred & Synth features to real terrains &  S2R & 25 \\
Features (2)& Real features to synth terrains &  R2S & 25 \\ \hline
\end{tabular}
\end{center}
\caption{Terrain type (real/synthetic/transferred features), categories, abbreviations, and the number of terrain samples in each category.}\label{tab:terrain-numbers}
\end{table}

\subsection{Rendering}
All terrains were rendered by using the same settings to avoid bias. The camera position was set to display the terrain from about~$45^o$ angle that is a common viewing distance from a top of a mountain or a low flying aircraft. This location shows enough details as opposed to a top view and does not cause self occlusions as opposed to a side view. The camera was positioned above one of the corners. We assumed viewers are familiar with this viewing angle.

We used sky sphere for illumination with gradient from~50\% of gray near the horizon to full white in zenith. The rendering was performed by using global illumination with no additional lights, by using 500 reflections and $9\times$ super-sampling for anti-aliasing. Each terrain was textured by the same color map that changed from low-level and flat areas with yellow color (sand), medium levels flat green (grass) to high and steep slopes gray (stone). We intentionally used non-photorealistic rendering~\cite{gooch1998non} so as to avoid any bias introduced by the simulation of vegetation and realistic rock, sand or grass rendering. Moreover, non-photorealistic rendering enhances the shape and structure of the bare elevation of terrain which is the focus or this study.

The image resolution used for the perceptual experiment was given by the size of the screen used in Mechanical Turk. The terrain DEM resolution was calculated by scaling down a terrain from $1,201\times{1},201$ that was the maximum available resolution for LIDAR scans by 10\% down to $128\times{128}$ and comparing the Peak Signal to Noise Ratio of the heightmaps and the rendered images. The error between the maximum resolution and $512^2$ was only 19.3\% and it provided a good compromise in terms of training deep neural networks, rendering, viewing image pairs without zooming in and out while testing.

\section{Perceptual Study and Feature Transfer}\label{sec:tests}
The perceptual study was run on the Amazon Mechanical Turk and we asked the subjects: \textit{"Which terrain looks more realistic (left or right)?"} by showing a pair of terrain images without giving any other information about the terrain. Each image pair was shown only once to each participant, but each image repeated several times in different pairs. The survey was blinded such that the participants only see an image pair with responses restricted to 'Left' or 'Right' option. The experiment involved 70 participants with no particular constraints on their education or previous knowledge and all participants were older than 18 years. However, only qualified "Mechanical Turk Masters", \ie users who consistently demonstrate accuracy in answers, were allowed to answered the survey. 

For each image pair we denote the category by a dash, so R-S indicates pair of images where one is from the real and one from the synthetic sample. The actual position of each image (left or right) was randomized making this relation symmetrical: R-S is the same as S-R.

\subsection{Experiment 1: Real and Synthetic Terrains}\label{sec:test1}
We generated random image pairs by using the rendered images from Sec.~\ref{sec:data}. We randomly paired one real with one synthetic image resulting in 150 image pairs. This pairing happened five times for each image from R resulting in $|R-S|=750$ images.

We made sure that the pairing did not miss any image, each image was repeated exactly five times, and pairing occurred always with a different image. The order of the images within each pair was randomized so that the synthetic image could be on the left hand or right hand side of the pair with the same probability. 

Each image pair was shown to five different participants resulting in a total 3,750 image pair observations by a total of 70 subjects with varying degree of participation (determined based on the unique count of anonymized 'workerID' provided by Amazon Mechanical Turk). 

Each time an image was selected as more realistic, it received a point, and the total number of points determined the overall ranking of each image that was normalized (see Sec.~\ref{sec:results}). Moreover, we also calculated the normalized ranking of each category of real (RA, RC, RG, RF, and RS) and synthetics (SP, SR, SM, ST, SF, and SC) terrains.

\subsection{Feature Transfer}\label{sec:gan}
E1 provided ranking of each category of real and synthetic terrains and the distribution of geomorphons confirmed (see Sec.~\ref{sec:geomres}) that they are related to the perceived quality. High-ranking real terrains contained features such as the valley topology in the terrains with fluvial erosion that were almost absent in low-ranking ones.  

We assumed that a deep neural network could learn the features that make real terrains visually plausible and that such features can be transferred onto the synthetic terrains to make them more visually plausible. Similarly, we hypothesize that the transfer could diminish features if it occurs from synthetic to real terrains that would further justify the importance of specific features for perceived realism.  

\begin{figure}[hbt]
  \centering
  \includegraphics[width=\linewidth]{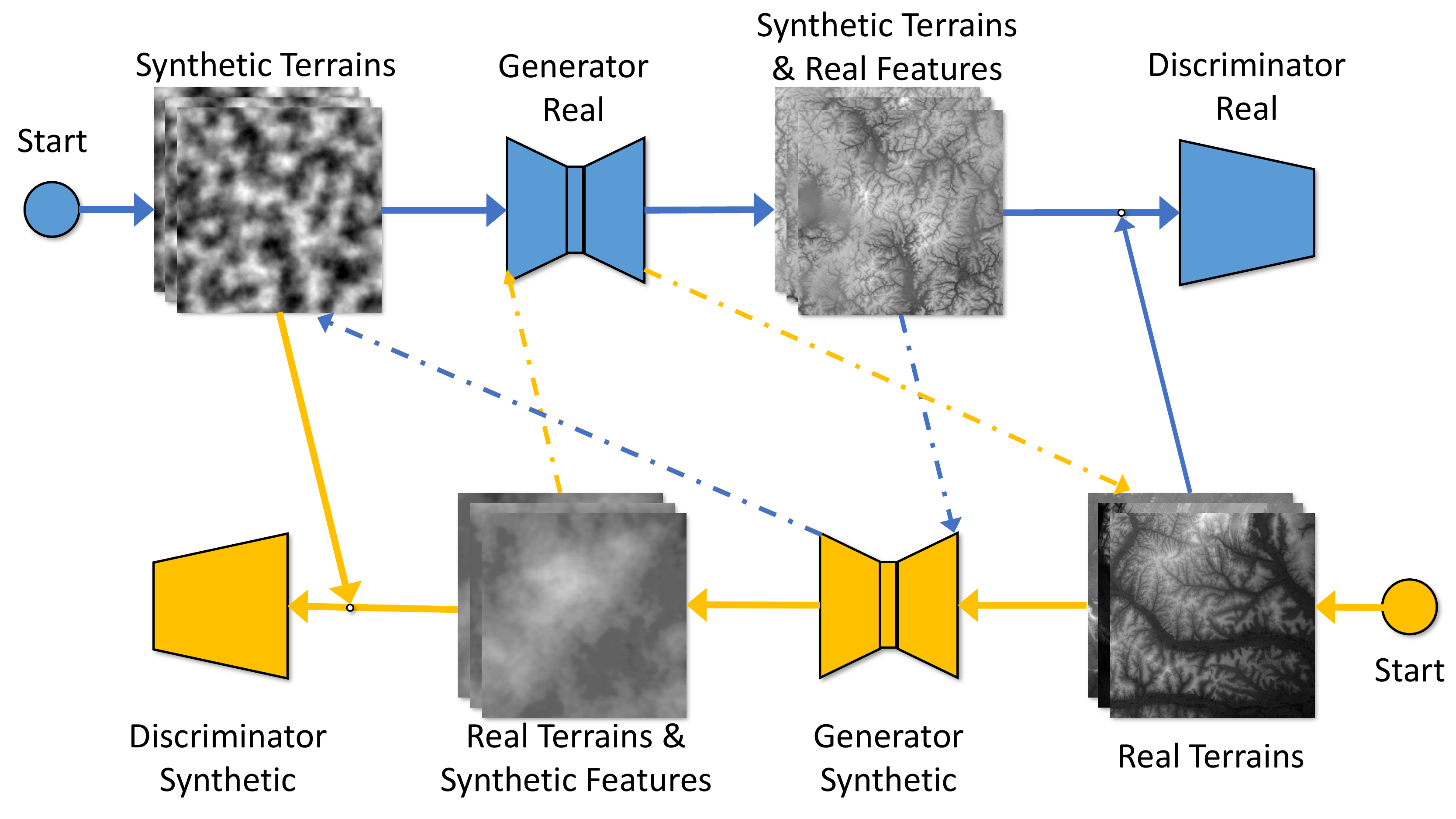}
  \caption{Feature transfer: The blue arrows indicate the working flow of R2S; the orange arrows indicate the working flow of S2R. The dotted-and-dashed arrows indicate the cycle consistency process.}
  \label{fig:gan}
\end{figure}

Because the explicit pairing between the real and synthetic terrains is difficult, we used the unpaired image to image translation~\cite{zhu:2017:ICCV} to transfer features from the real domain to the synthetic domain, and vice versa (see Fig.~\ref{fig:gan}). We use a pair of generators~$G_R$ and~$G_S$ with a pair of discriminators~$D_R$ and~$D_S$. The generator~$G_R$ translates terrains from the synthetic domain~$S$ to the real domain~$R$ with real features. The discriminator~$D_R$ discriminates between terrains~$\{r\}$ and~$\{G_R(s)\}$, where $\{r\}\in~R$ and~$\{s\}\in~S$. Moreover, $G_S$ translates terrains within the real domain~$R$ to the synthetic domain $S$ with synthetic features. Similarly, $D_S$ discriminates between terrains~$\{s\}$ and~$\{G_S(r)\}$, where~$\{r\}\in~R$ and~$\{s\}\in~S$. Besides the adversarial loss, a cycle consistency loss is used to make $G_S\left(G_R(s)\right)\approx{s}$ and~$G_R\left(G_S(r)\right)\approx{r}$. This process is indicated with the dashed arrows in Fig.~\ref{fig:gan}. The cycle consistency ensures the high-quality feature transfer.

We adopt a nine res-block generator and a $70\times{70}$ PatchGAN discriminator~\cite{isola:2016:ARXIV}. The transfer generated checkerboard patterns 
caused by fractionally-strided convolution and the artifacts decrease if the training epochs increase. We also applied resize-conv with Nearest Neighbor and Bilinear as suggested in~\cite{odena:2016:DISTILL}. 

Our training set contains $9,800$ real terrain height maps selected from the SRTM DEMs excluding the terrains that have been used in E1 and E2. We generated additional synthetic height maps for use in training based on aforementioned synthetic categorization and same size as the real terrain training data which is $9,800$ (see the data collection in Sections~\ref{sec:terrain_data_real} and~\ref{sec:terrain_data_synthetic}). 

\begin{figure}[hbt]
  \centering
  a)\includegraphics[width=0.45\linewidth]{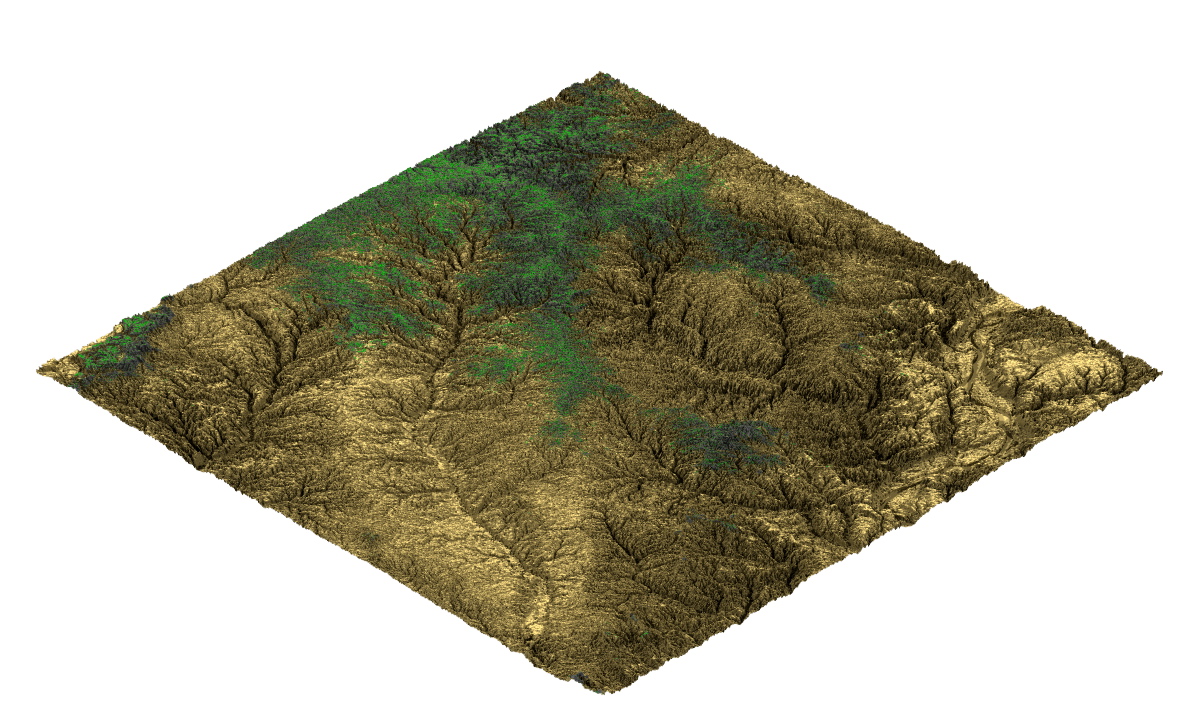}
  b)\includegraphics[width=0.45\linewidth]{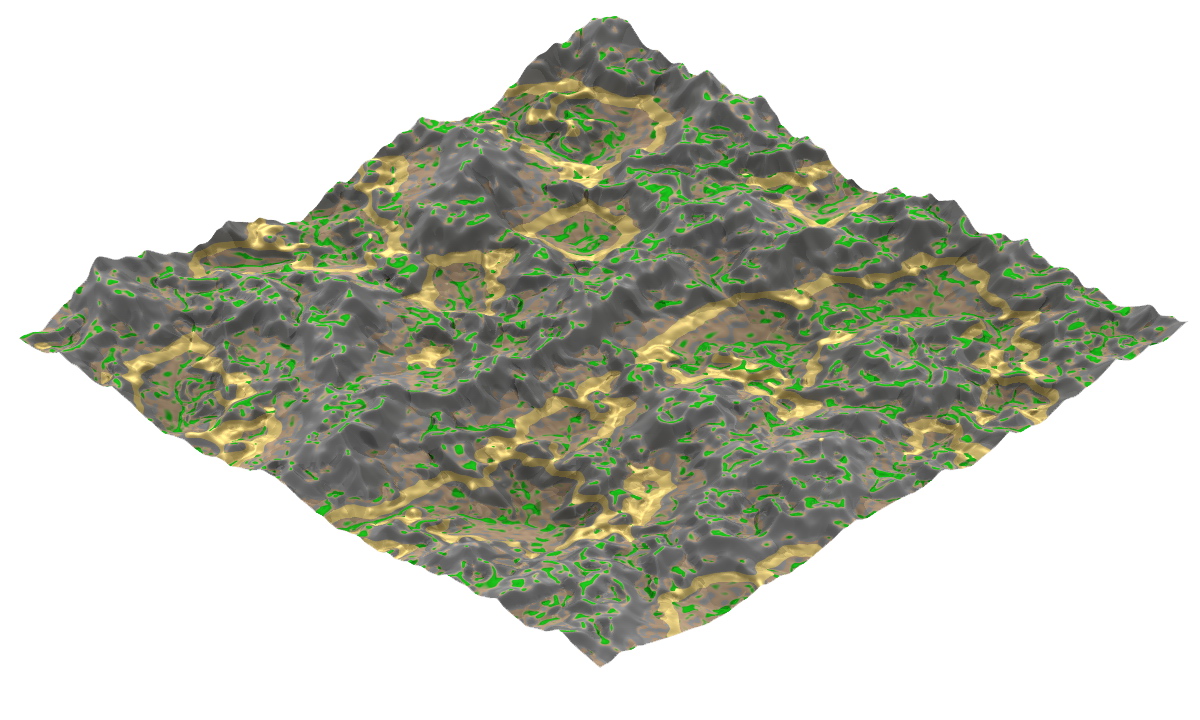}
  c)\includegraphics[width=0.45\linewidth]{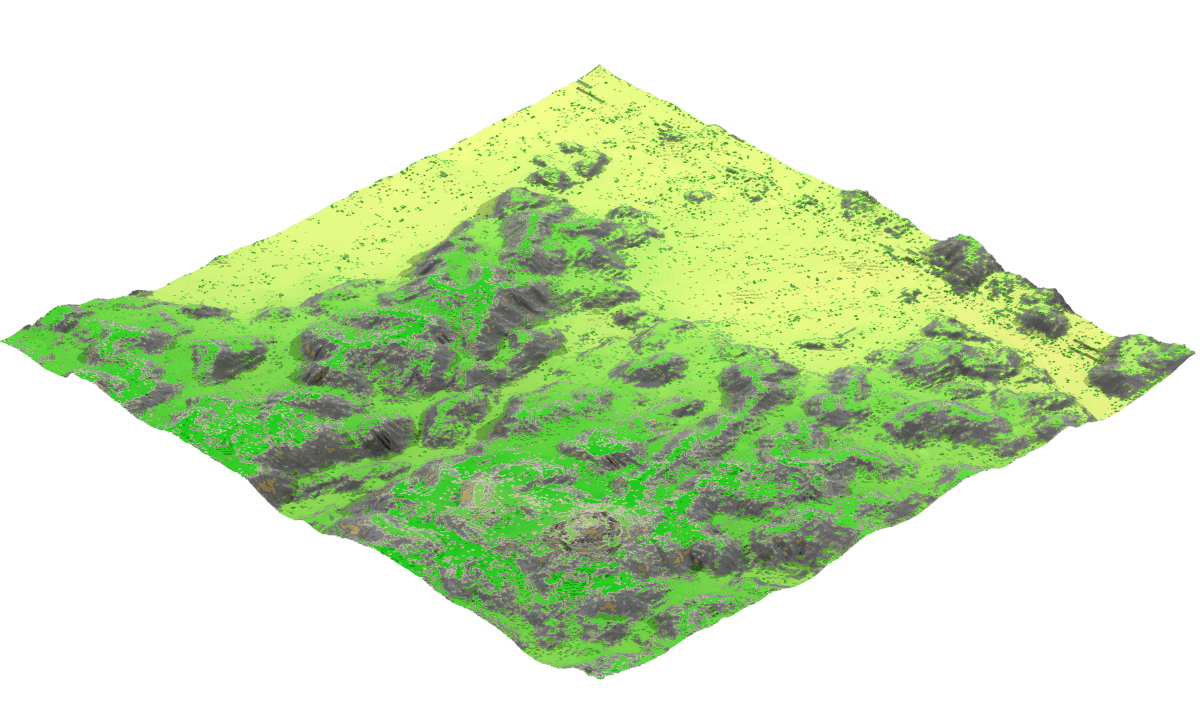}
  d)\includegraphics[width=0.45\linewidth]{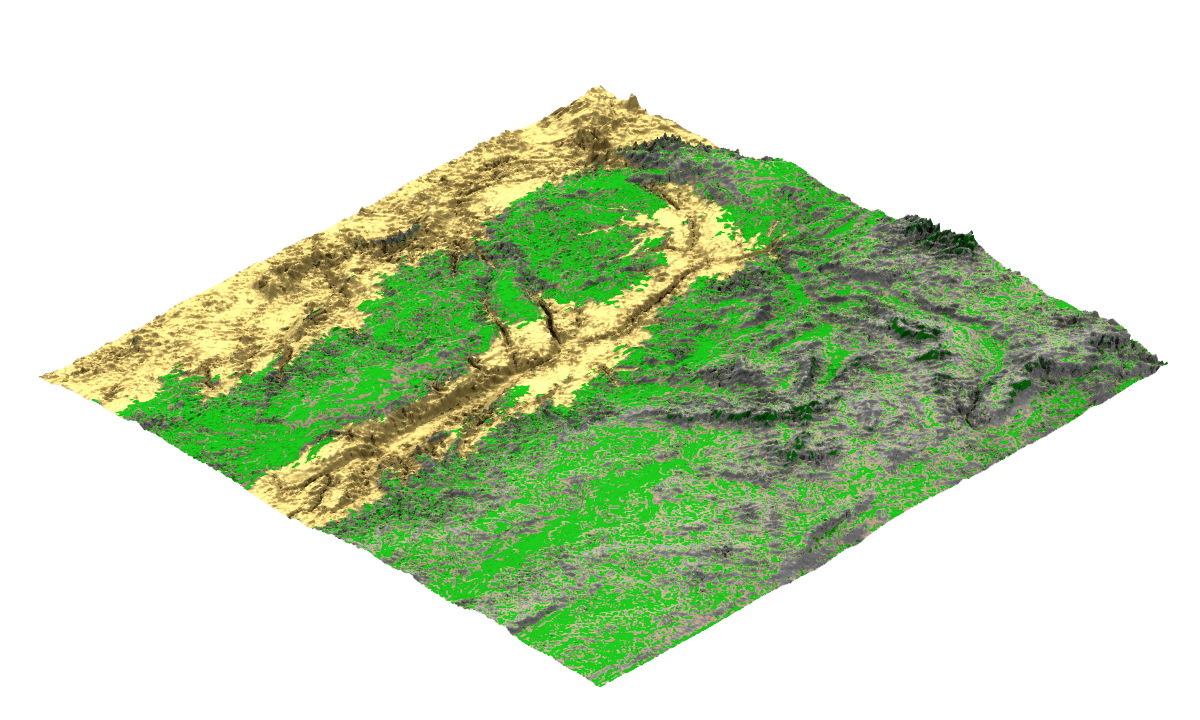}
  \caption{Example of feature transfer: a)~Real terrain with strong fluvial patterns from Colombian Amazonian forest area (S01 W072) (PTRM=0.67) and b) synthetic terrain generated by thermal erosion (PTRM=0.46). c)~Synthetic features transferred to real terrain worsen its perceived visual quality (PTRM=0.49) and d)~real features transferred to synthetic terrain improve it (PTRM=0.63).}
  \label{fig:transfer}
\end{figure}

We trained the model with $20$ epochs, and then generated 150 images of real terrains with synthetic features denoted by S2R. The term S2T denotes the \textit{transfer}, meaning "synthetic features were transferred to real terrains". We also generated another 150 images of synthetic terrains with real features denoted by R2S. Figures~\ref{fig:teaser} and~\ref{fig:transfer} show example result of the feature transfer in both directions (from real to synthetic and from synthetic to real) and Sec.~\ref{sec:results} further discusses results.

\subsection{Experiment 2: Real, Synthetic, and Terrain Models with Transferred Features}\label{sec:test2}
The objective of the second experiment (E2) was to evaluate how the terrains with transferred features score perceptually against real and synthetic terrains. We have reused the 750 R-S image pairs from E1 (Sec.~\ref{sec:test1}) and added another 750 images for each missing combination. Tab.~\ref{tab:combinations} shows the naming of the image pairs. The first column shows the reused pairs from E1 (R-S). The newly added pairs compare newly created transferred features from synthetic to real R2S combined with all options R2S-R, R2S-S, and S2R-R2S. Also, we added combinations for feature transfer from real to synthetic S2R \ie S2R-R and S2R-S. R2S-S2R is already included because it is symmetrical with S2R-R2S. 
\begin{table}[hbt]
\begin{center}
\begin{tabular}{|c||c|c|c|}
\hline
    & S            & R2S        & S2R \\ \hline \hline
R   & R-S          & R2S-R      & S2R-R\\ \hline
S   & $\bullet$    & R2S-S      & S2R-S\\ \hline 
R2S & $\bullet$    & $\bullet$  & S2R-R2S\\ \hline 
\end{tabular}
\end{center}
\caption{Image pairing for Experiment 2. R-S pairs are reused from E1.}\label{tab:combinations}
\end{table} 

As in E1, each shuffling was generated five times resulting in $750$ images for each item of Tab.~\ref{tab:combinations} resulting in total of $4,500$ image pairs. We have repeated each test for five independent viewers and this resulted in the total of $22,500$ views by $128$ subjects. All participants were older than 18 years and we again used only qualified Mechanical Turk Masters. Note that because the R-S set from the first experiment were also included in the second one, we have validated the first experiment, because the ranking of the results was consistent between E1 and E2 suggesting the data saturation point has been attained (Sec.~\ref{sec:results}).

\section{Results}\label{sec:results}
Here we discuss results of our two experiments and feature transfer. We show results of E1 and E2, discuss the features in geomorphons, and the feature transfer. Finally, we introduce the perceptual terrain quality metrics  PTRM.

\begin{figure}[hbt]
  \centering
  \includegraphics[width=\linewidth]{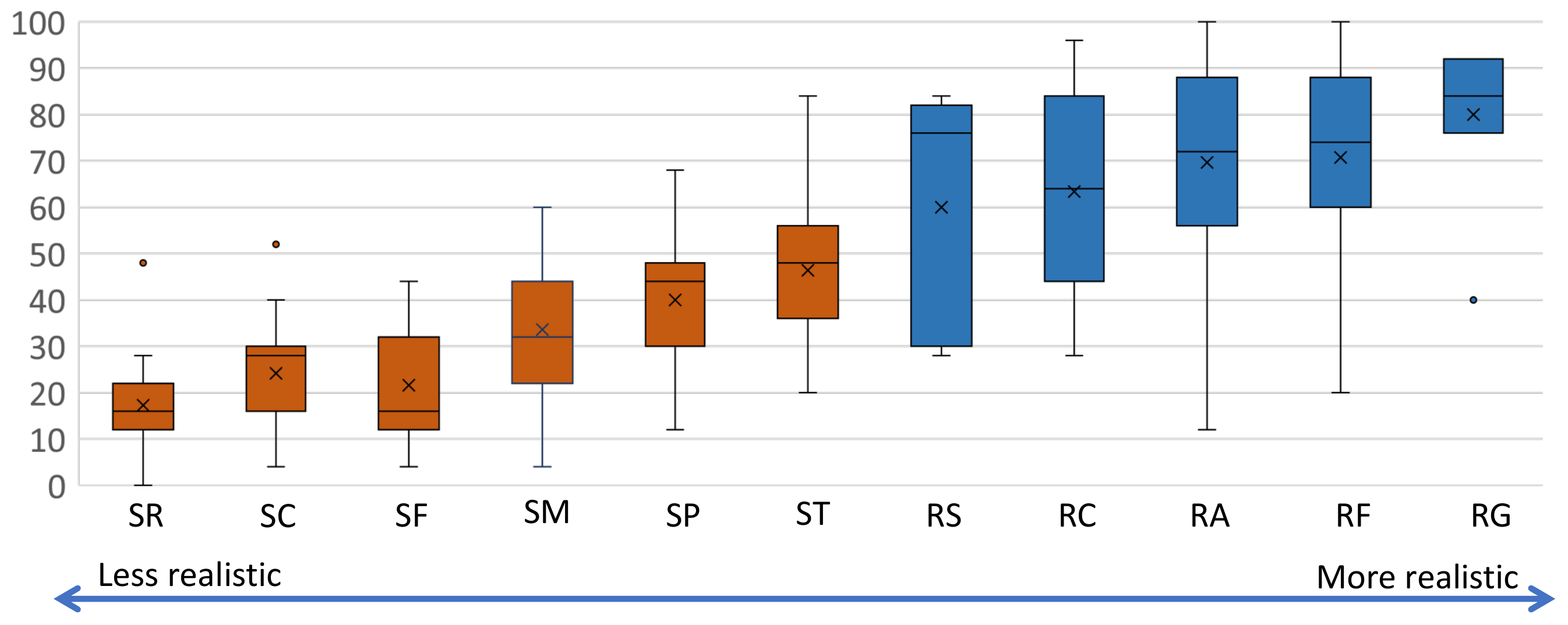}
  \includegraphics[width=\linewidth]{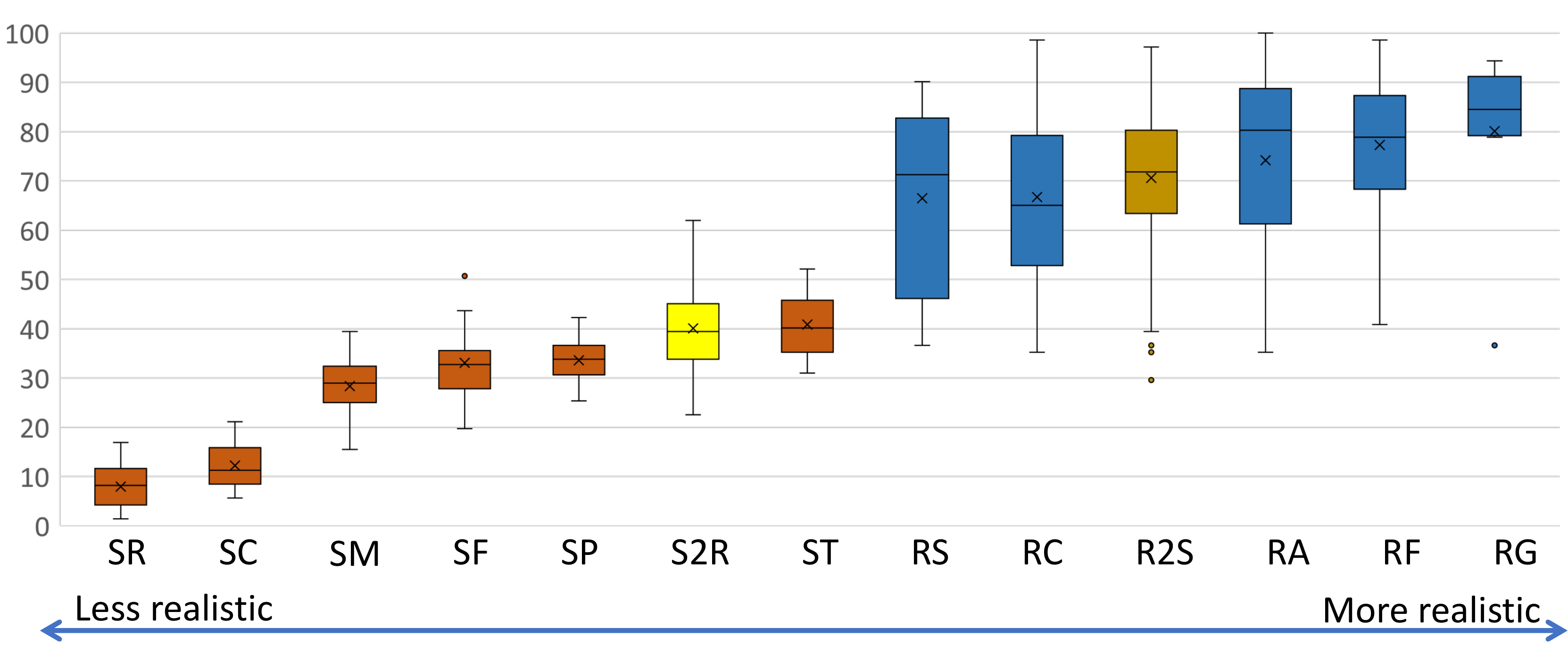}
  \caption{Perceptual ranking of terrains from E1 (top) and E2 (bottom). The abbreviations are from Tab.~\ref{tab:terrain-numbers} and the terrains are sorted by the average perceived realism from worse (left) to the best (right).
  While the order of the rankings in E2 is very similar to E1, note that the S2R \ie synthetic terrains improved with features from real terrains ranked high. At the same time, real terrain with features transferred from procedural R2S ranked lower. The figure has been plotted based on their average scores. The $\times$, $\bullet$, and the $-$ sign represent the mean, outlier points, and the median markers respectively. 
  }
  \label{fig:test_boxwhisker}
\end{figure}

\subsection{Perceptual Experiments}
\label{sec:perceptual_exp}
\textbf{Experiment 1} assigned each image a number of how many times it was selected as more realistic in a pair-wise choice randomized test. We normalized the counts so that the most realistically perceived image had a score of 1.0. We then calculated the average, standard deviation, mean, and range for each category of R and S from Tab.~\ref{tab:terrain-numbers}. The sorted results by the average value are shown in Fig.~\ref{fig:test_boxwhisker} top. The ranking of terrains from least realistic to the best was: SR-SC-SF-SM-SP-ST-RS-RC-RA-RF-RG. All synthetic terrains were perceived as visually less realistic than the real ones. The most realistic synthetic terrains were generated by thermal erosion (ST) (see Tab.~\ref{tab:ex-results}).

We have also calculated the average and standard deviation of values of ranking of all images in the sets S and R. An unpaired T-Test evaluation suggested that the difference is statistically significant with the two-tailed ~$p<0.01$, $DF =283, t = 17.91 \& \alpha = 0.01$.

The perceptual experiment suggests that synthetic terrains in our data-set are perceived as visually significantly less realistic than the real ones.

\textbf{Experiment 2} repeated E1 with the addition of pairs of images with transferred features (Sec.~\ref{sec:gan}). Our assumption was that the features transferred from real terrains to synthetic would improve their ranking and that the transfer of features from synthetic to real terrains would do the opposite. The normalized rank of each image and calculated statistics for each category are in Fig.~\ref{fig:test_boxwhisker} and Tab.~\ref{tab:ex-results}.

The perceived order of terrain categories is the same as in E1 that confirms the validity of both tests. The categories with transferred features ranked as expected: the synthetic terrains enhanced with features from real terrains R2S ranked 10th, which is better than some of the real terrains (RS and RC), but better than all synthetic ones. This confirms our hypothesis that feature transfer affects terrain perception. Similarly, the real terrain with transferred procedural features S2R ranked significantly worse than real terrains and even worse than thermal erosion simulation (ST) at 6th place. This confirmed our hypothesis that features of synthetic terrains do not contribute significantly to terrain realism.

\subsection{Statistical Tests}
We performed statistical tests on our normalized perceptual scores to determine if there are any differences in perception of our terrain data groups: R, S, R2S and S2R. We state the null hypothesis, H\textsubscript{0} for our six statistical tests in E2 as follows: \textit{``There are no significant differences in the visual perception scores between our terrain data groups.''}.

We used T-Test to compare the means and variances of the perception scores and the results are summarized in Tab.~\ref{tab:ex-results}. For testing our candidates in E2, we used the significance level of $\alpha=0.01$, and get the statistics for, R versus R2S ($p=0.02$, $DF =149, t = 2.26$), R versus S2R ($p<0.01$, $DF =149, t = 22.10$), R versus S ($p<0.01$, $DF =149, t = 22.59$), R2S versus S2R ($p<0.01$, $DF =149, t = -23.52$), R2S versus S ($p<0.01$, $DF =149, t = 29.12$) and S2R versus S ($p<0.01$, $DF =149, t = 10.79$). Tab.~\ref{tab:ex-stat_significance} summarizes the perception scores. The scores are statistically different between the terrain groups. The observers perceived the realism of the terrains at different scales except the R vs R2S. This implies that there are features in real terrains that increase the perceived realism and we can reject our null hypothesis stating that \textit{there is a significant difference in perception of Real Terrains (R), Synthetic Terrains (S), Synthetic Terrains with Real features (R2S), and Real Terrains with Synthetic features (S2R)}.
\begin{table}[hbt]
\begin{center}
\begin{tabular}{|l|l|l|l|l|}
\hline
    & R             & S             & R2S           & S2R           \\    \hline
R   & $\bullet$     & $\checkmark$  & $\times$  & $\checkmark$  \\    \hline
R2S & $\bullet$     & $\bullet$     & $\checkmark$  & $\checkmark$  \\    \hline
S2R & $\bullet$     & $\bullet$     & $\bullet$     & $\checkmark$  \\    \hline
S   & $\bullet$     & $\bullet$     & $\bullet$     & $\bullet$     \\    \hline
\end{tabular}
\end{center}
\caption{The table shows the statistical significance of each terrain set compared with the other set from our experiments: E1 and E2. The $\checkmark$ implies that the terrain set in the vertical column are statistically significant than the terrain set in the horizontal row, $\times$ to suggest that the difference is not statistically significant and $\bullet$ to suggest that the test is not available or compared already. }\label{tab:ex-stat_significance}
\end{table}

\begin{table*}[hbt]
\begin{center}
\begin{tabular}{ |l| l|| c| c| c| c| c| c| c|| c| c| c| c| c| c| c|}
 \hline
  \multicolumn{2}{|c||}{}  & \multicolumn{7}{c||}{E1}      & \multicolumn{7}{c|}{E2}\\ \hline
 \textbf{T} & \textbf{Ab.} & \textbf{AVG} & \textbf{MED} & \textbf{MODE} & \textbf{RNG} & \textbf{STDEV} & \textbf{SE} & \textbf{95\% C.I.} & \textbf{AVG} & \textbf{MED} & \textbf{MODE} & \textbf{RNG} & \textbf{STDEV} & \textbf{SE} & \textbf{95\% C.I.}\\ \hline
 R  & RG  & 80 & 84  & 92 & 52 & 19 & 7  & 14  & 80 & 851  & N/A & 58 & 19 & 7 & 13\\
    & RF  & 71 & 74  & 88 & 80 & 20 & 3  & 5   & 77 & 79  & 86 & 58 & 13 & 2 & 3\\
    & RA  & 70 & 72  & 92 & 88 & 22 & 3  & 6   & 74 & 80  & 89 & 65 & 19 & 3 & 5\\
    & RC  & 63 & 64  & 96 & 68 & 21 & 5  & 9   & 67 & 65  & 63 & 63 & 17 & 4 & 8 \\
    & RS  & 60 & 76  & N/A& 56 & 28 & 12 & 24  & 67 & 71  & N/A & 53& 20& 8 & 16\\ \hline
S   & ST  & 46 & 48  & 48 & 64 & 17 & 3  & 7   & 41 & 40  & 32 & 21 & 7  & 1 & 3  \\
    & SP  & 40 & 44  & 48 & 56 & 13 & 3  & 5   & 34 & 34  & 34 & 17 & 5  & 1 & 2\\
    & SM  & 34 & 32  & 36 & 56 & 14 & 3  & 6   & 28 & 29  & 32 & 24 & 6  & 1 & 2\\
    & SF  & 22 & 16  & 16 & 40 & 12 & 2  & 5   & 33 & 33  & 35 & 31 & 8  & 2 & 3\\
    & SC  & 24 & 28  & 28 & 48 & 11 & 2  & 4   & 12 & 11  & 10  & 16 & 5  & 1 & 2 \\
    & SR  & 17 & 16  & 20 & 48 & 10 & 2  & 4   & 8  & 8   & 13 & 16 & 4  & 1 & 2 \\  \hline
2   & R2S & N/A   & N/A    & N/A   & N/A   & N/A   & N/A   & N/A    & 71 & 72  & 70 & 68 & 13 & 2 & 2\\
    & S2R & N/A   & N/A    & N/A   & N/A   & N/A   & N/A   & N/A    & 40 & 39  & 33 & 39 & 9  & 1 & 1    \\   \hline
\end{tabular}
\end{center}
\caption{The Average (AVG), Median (MED), Mode (MODE), RANGE (RNG), Standard Deviation (STDEV), Standard Error (SE), and 95\% Confidence Interval (95\% C.I.) of the normalized scores for the terrain sets: E1 and E2.}\label{tab:ex-results}
\end{table*}

We performed an ANOVA (E1: $calculated~F = 320.91$, $critical~F = 3.87$, $p<0.01$, $df = 298$) (E2: $calculated~F = 465.78$, $critical~F = 2.61$, $p<0.01$, $df = 596$) to determine if there are any significant differences in the variances of the scores. After establishing that there are differences in the groups, we proceeded with the T-Tests to determine among which groups the significant differences lie. Additionally, a post-hoc test Tukey's Honestly Significant Difference (HSD) test indicated that there is no statistically significant difference in the perception scores between the terrain groups, R and R2S with a $p=0.0511$ and standard error of $1.0938$ while there is a statistically significant difference between the rest of the terrain groups with $p<0.001$ and standard error of 1.0938 with $\alpha=0.01$ which is consistent with T-Test results.

\subsection{Geomorphons}\label{sec:geomres}
\begin{figure}[hbt]
  \centering
  \includegraphics[width=\linewidth]{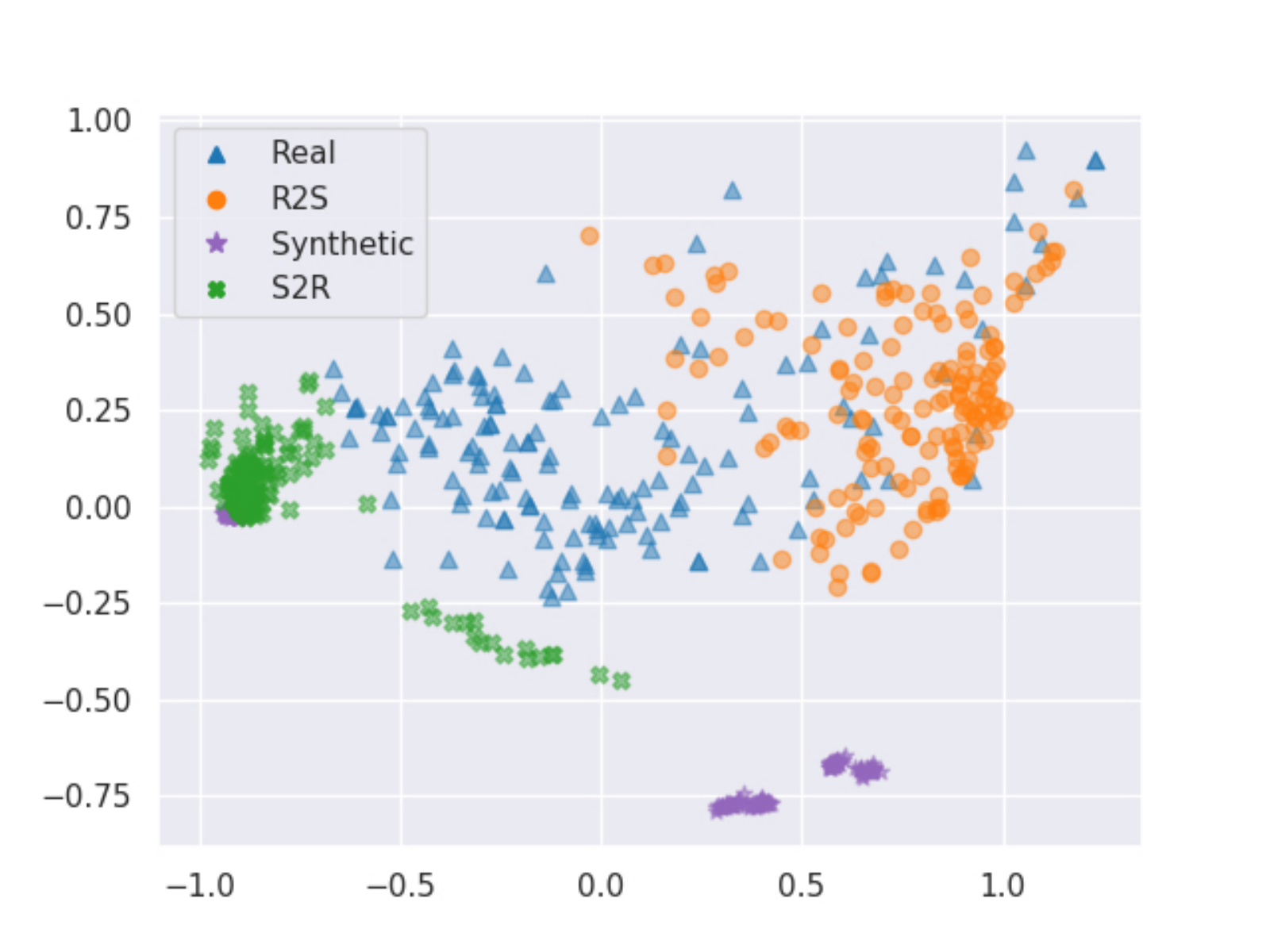}
  \caption{Projection of geomorphons from all terrains to 2D. Synthetic terrains appear clustered, while real terrains are more scattered. Transfer of real features scatters the terrains and transfer of procedural features cluster the resulting terrains.}
  \label{fig:clusters}
\end{figure}

Geomorphons (Sec.~\ref{sec:geomorphons}) characterize local terrain features (valleys, ridges, peaks, etc.). A~geomorphon is a 10D feature vector that describes a terrain. The spatial distribution of geomorphons brings further insight into the features and the corresponding data-sets. Fig.~\ref{fig:clusters} shows the points corresponding to all our data-sets (R, S, R2S, and S2R) projected from 10D space to 2D by using t-Distributed Stochastic Neighbor Embedding algorithm~\cite{maaten2008visualizing} that preserves distances among points across the dimensions.

\begin{figure*}[hbt]
\raggedright
 \includegraphics[width=0.3\linewidth]{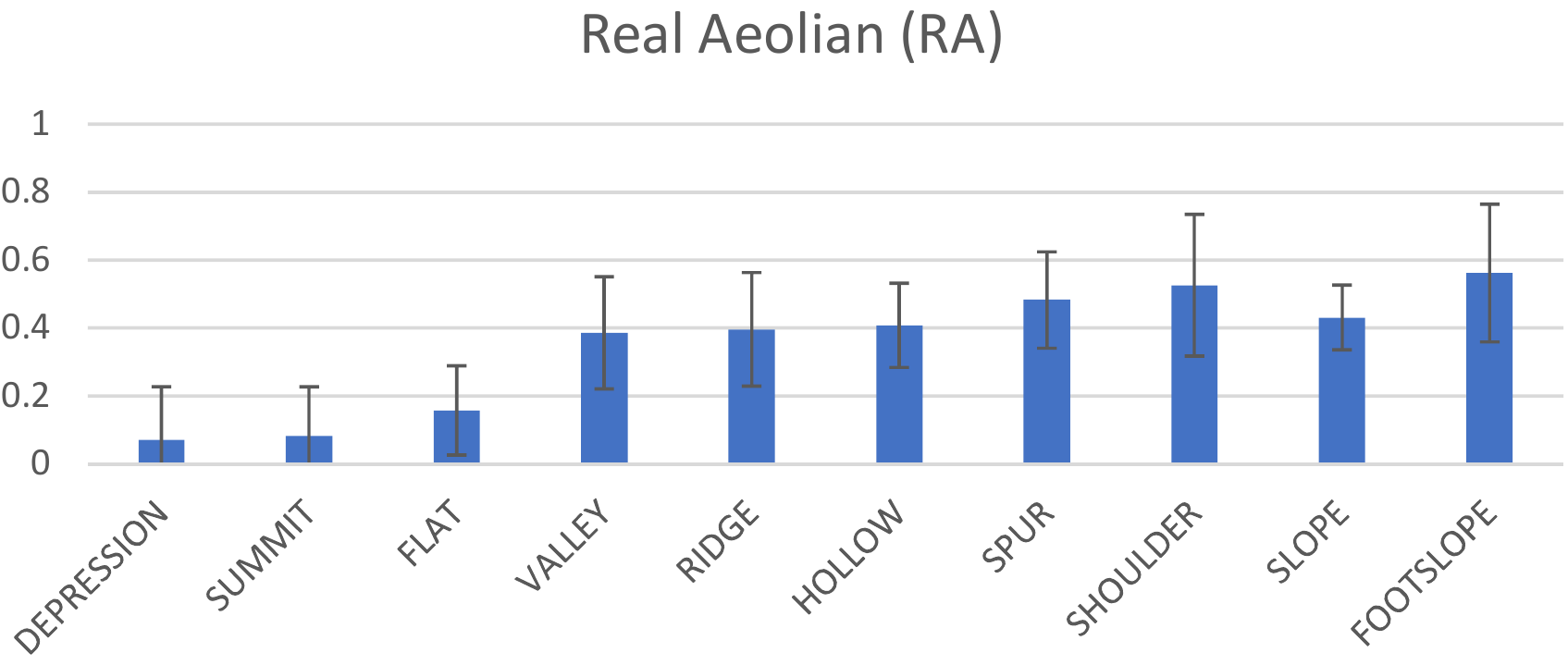}
 \includegraphics[width=0.3\linewidth]{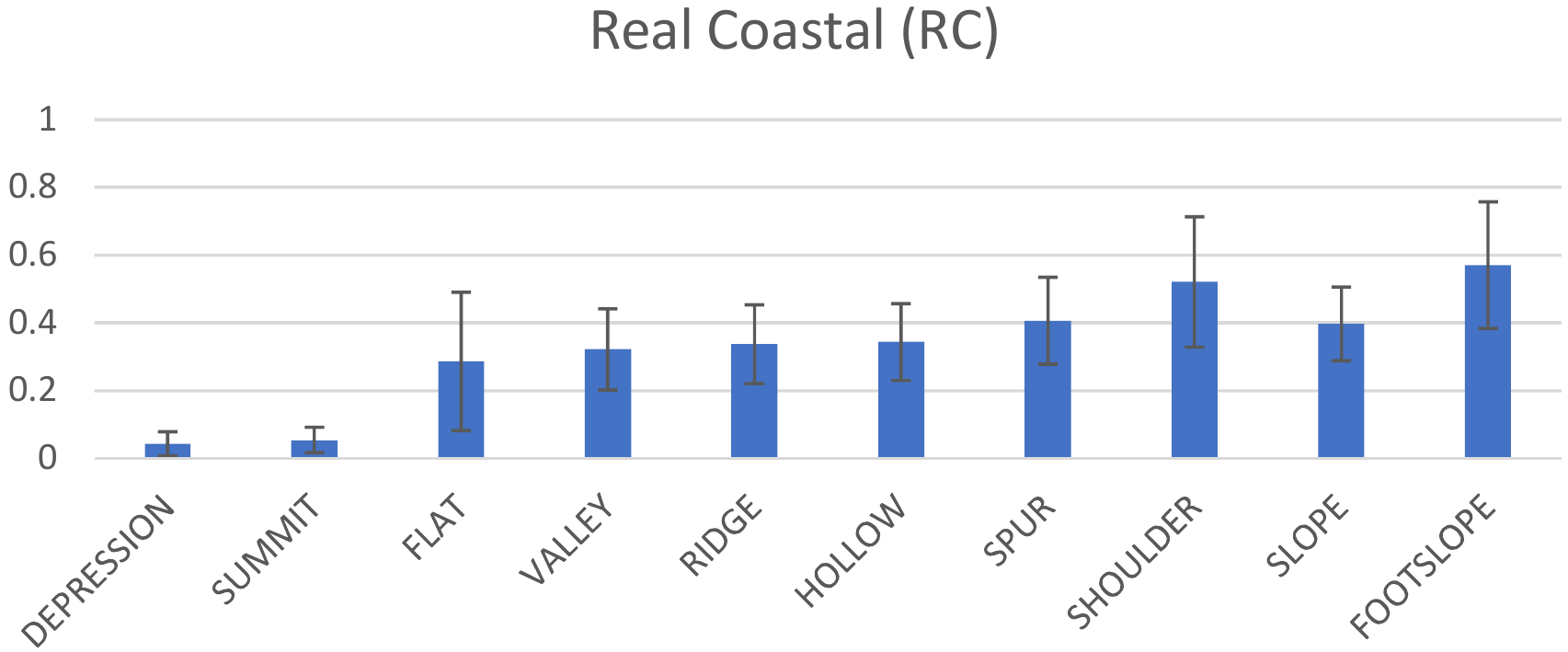}
 \includegraphics[width=0.3\linewidth]{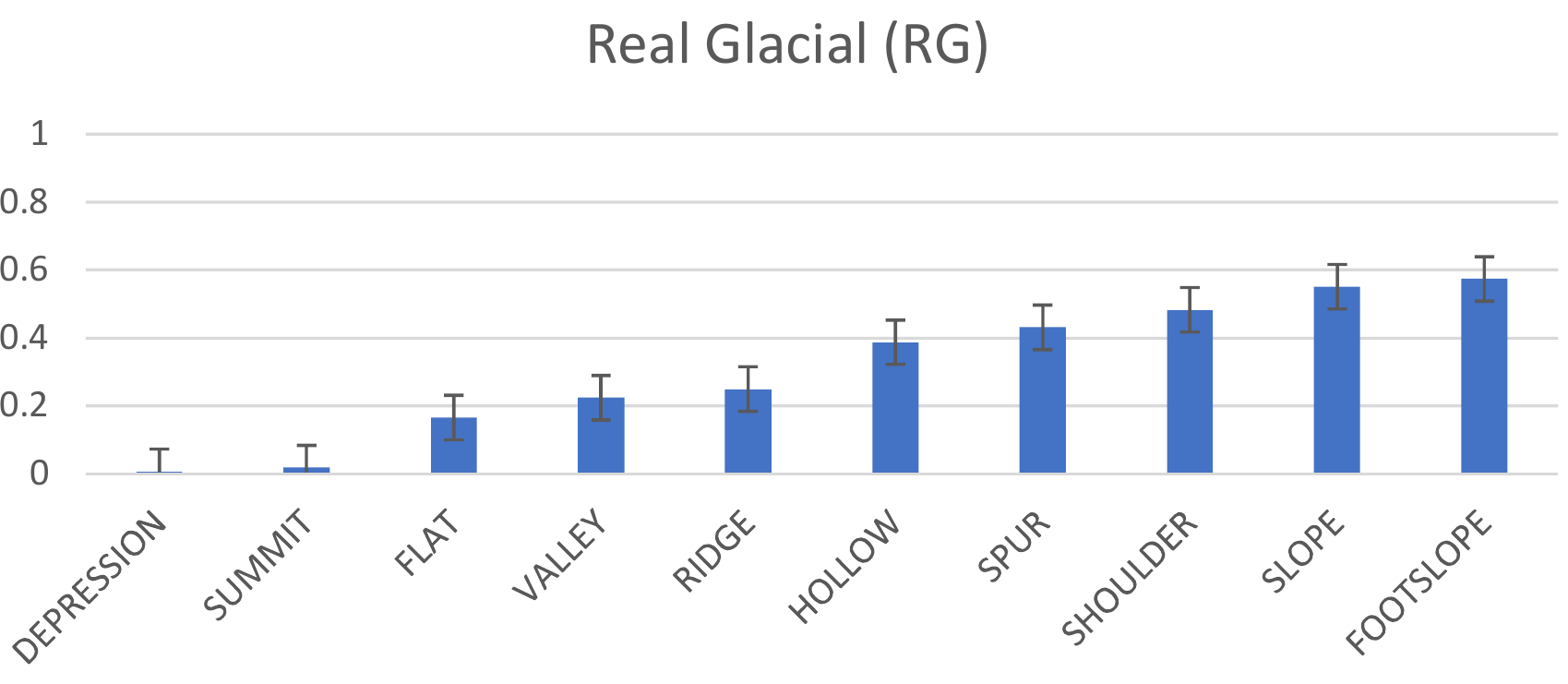}
 \includegraphics[width=0.3\linewidth]{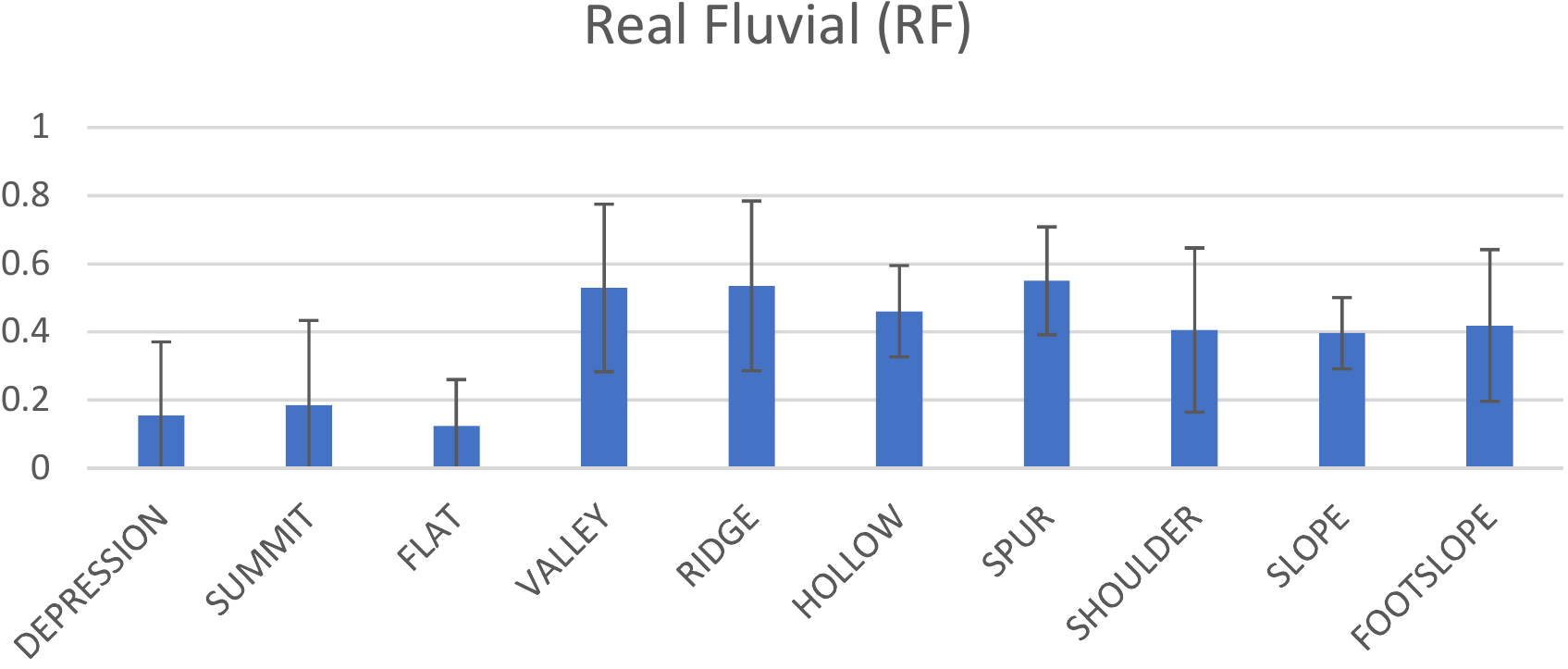}
 \includegraphics[width=0.3\linewidth]{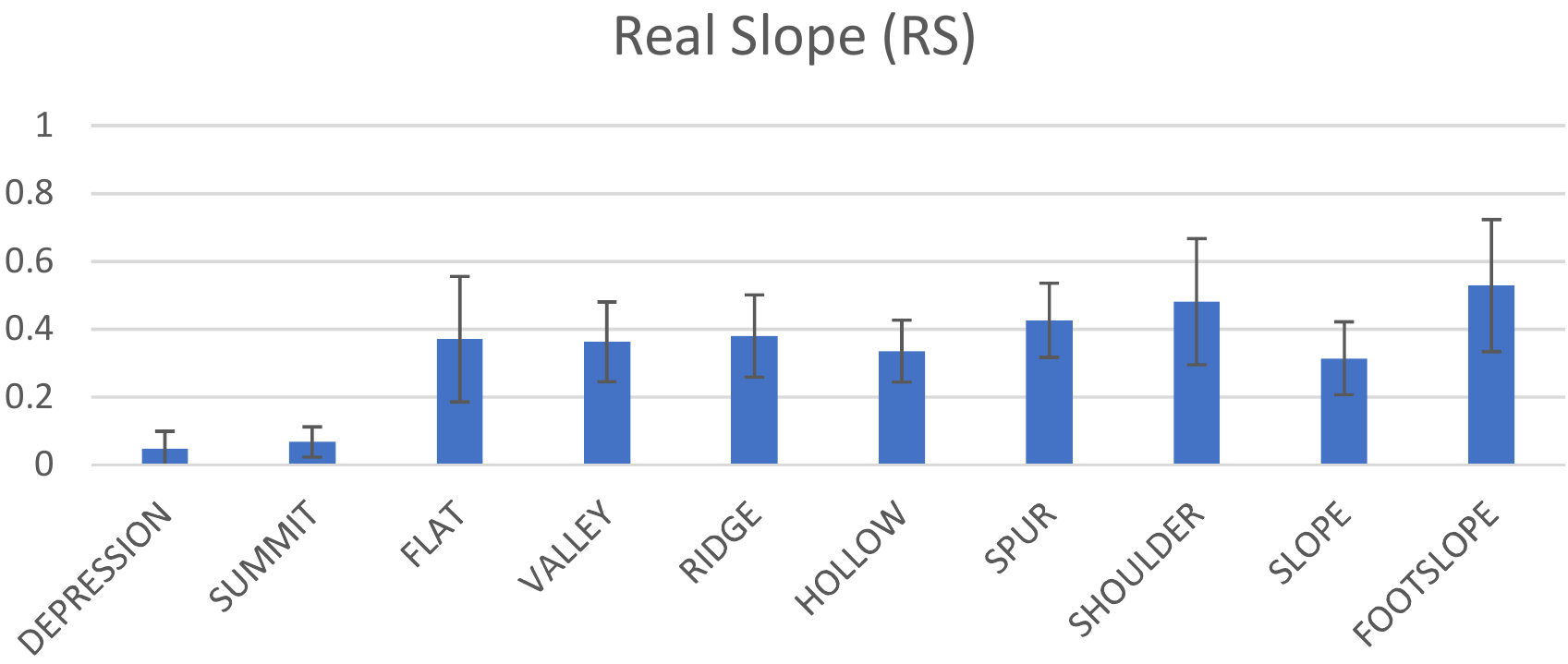}\\
 \includegraphics[width=0.3\linewidth]{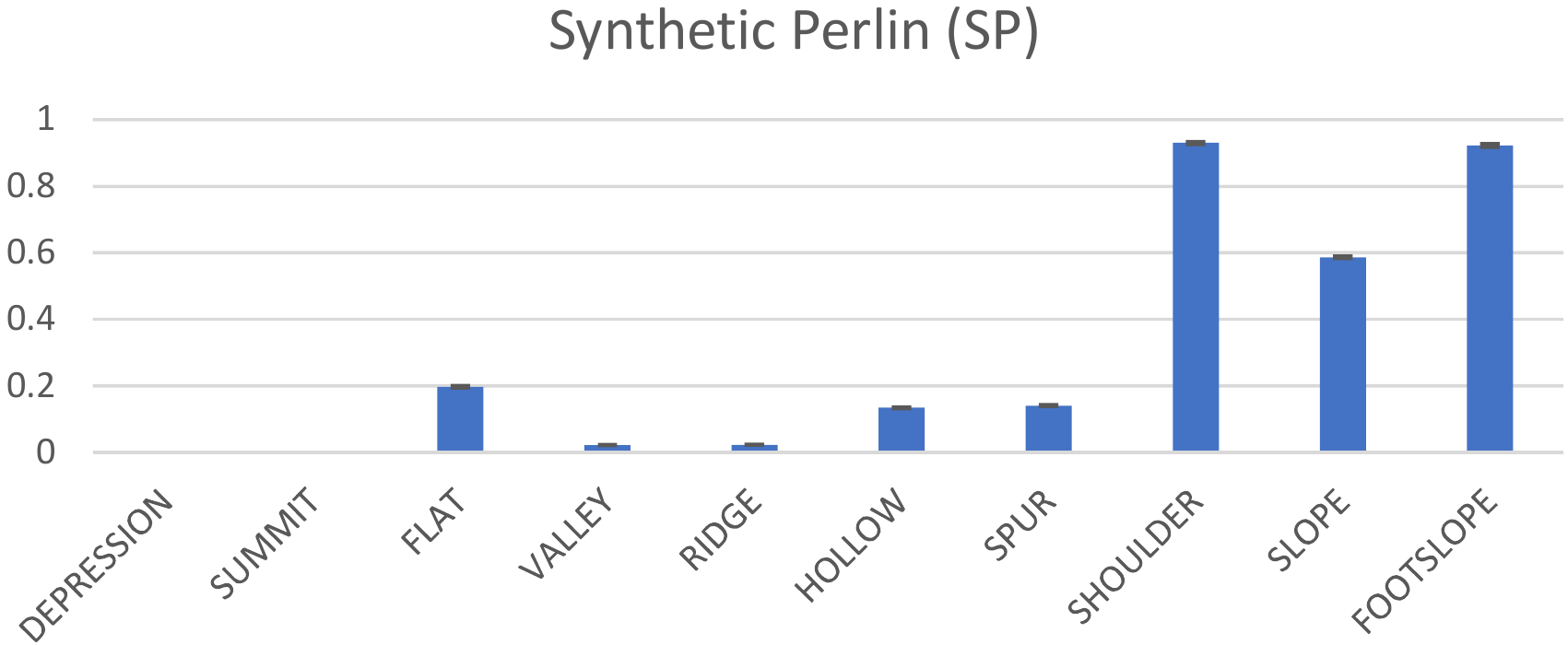}
 \includegraphics[width=0.3\linewidth]{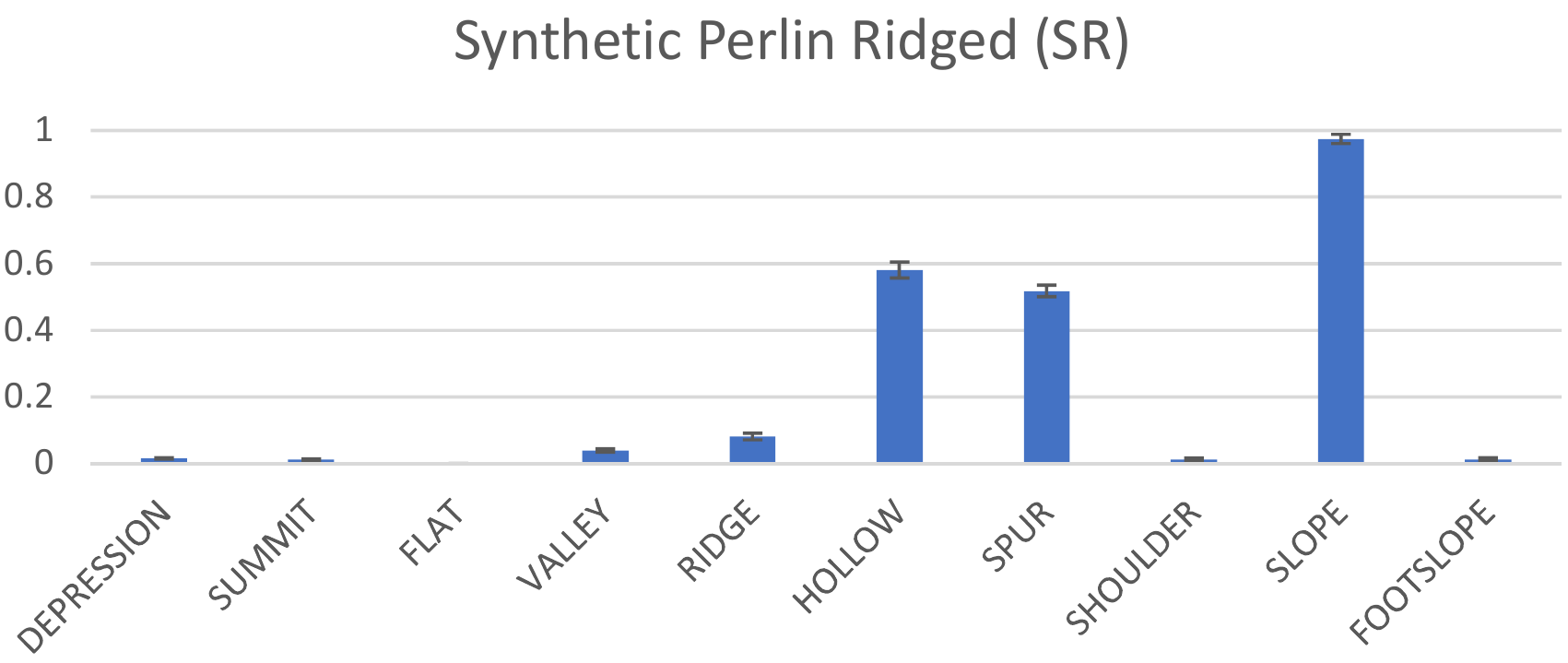}
 \includegraphics[width=0.3\linewidth]{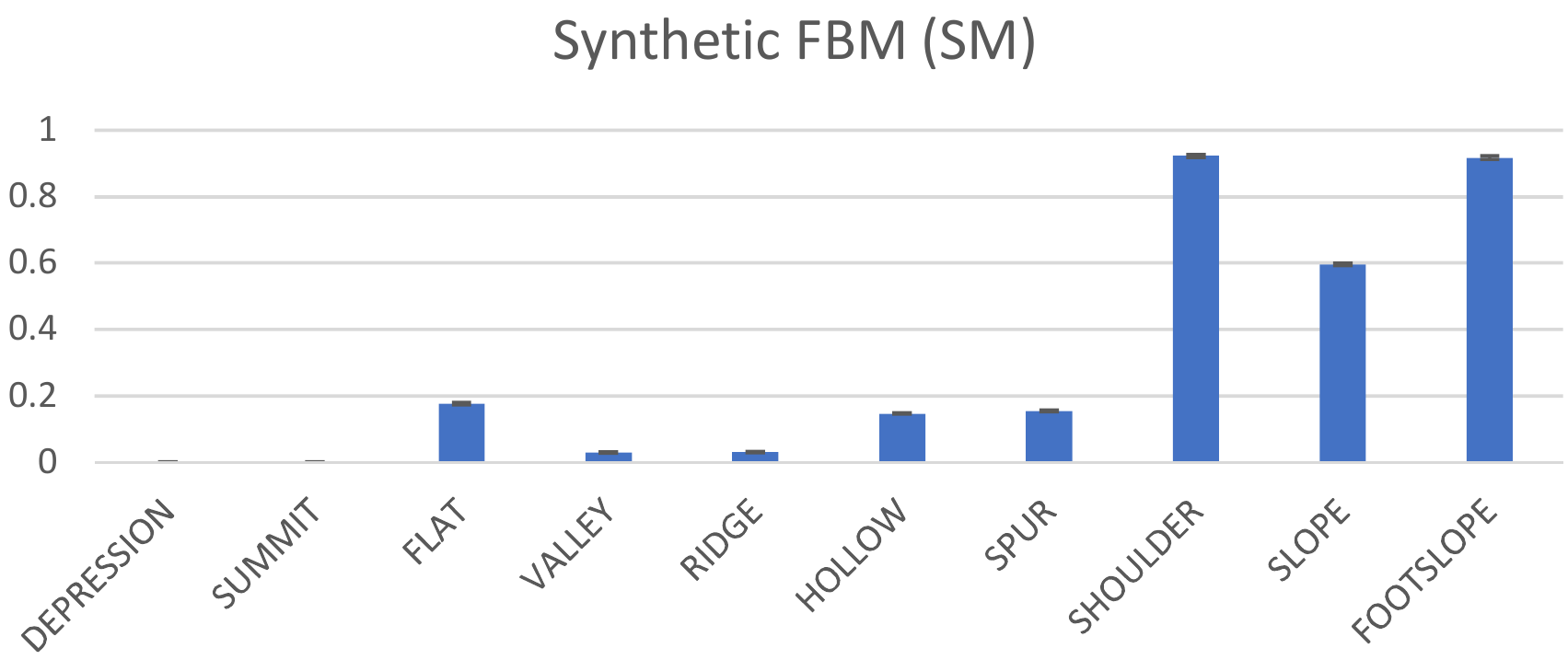}
 \includegraphics[width=0.3\linewidth]{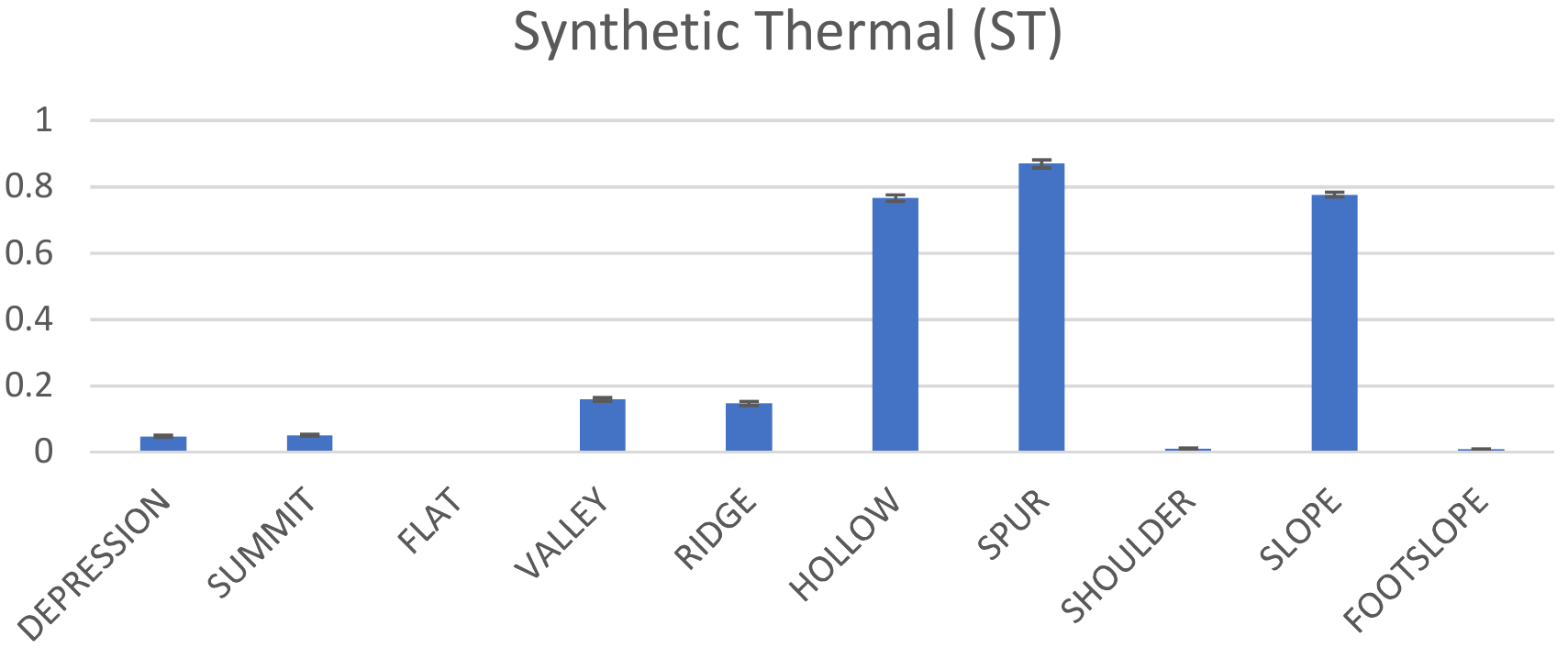}
 \includegraphics[width=0.3\linewidth]{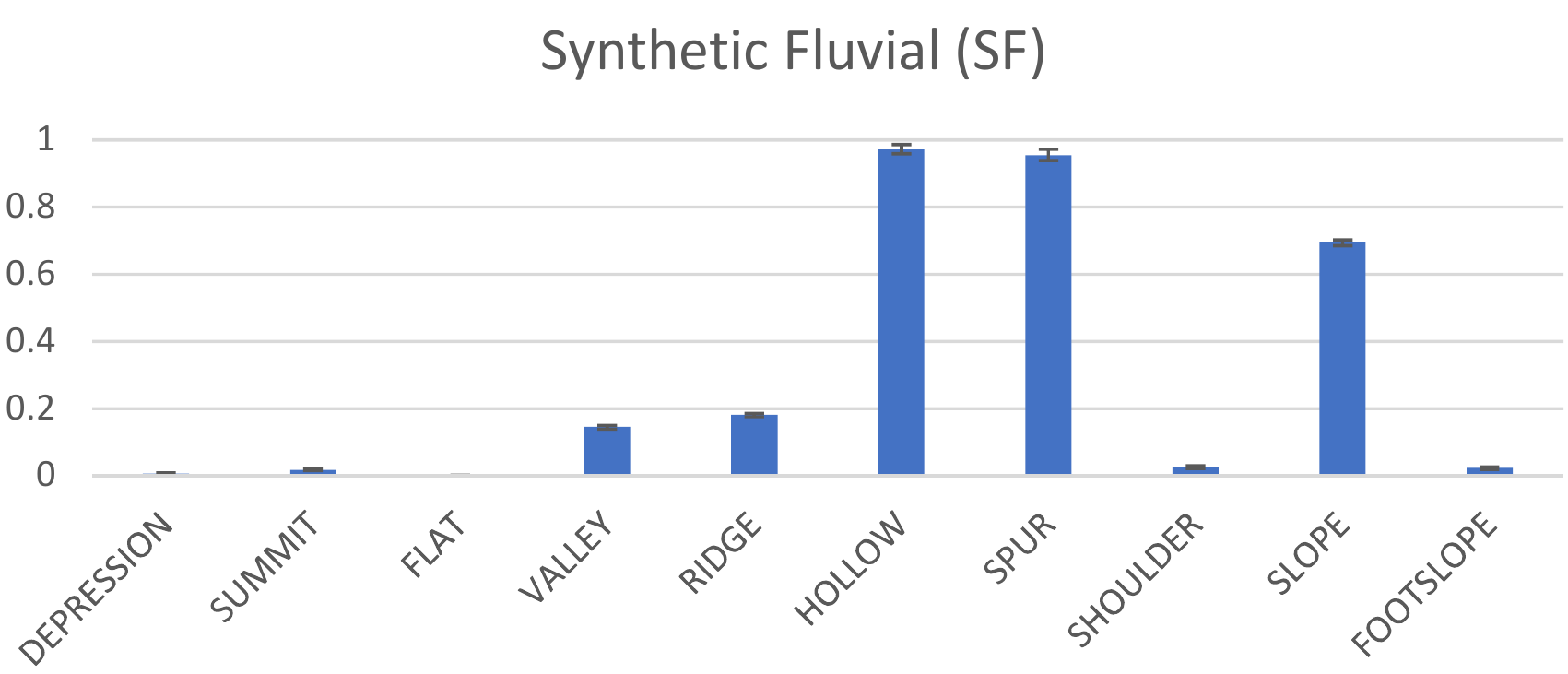}
 \includegraphics[width=0.3\linewidth]{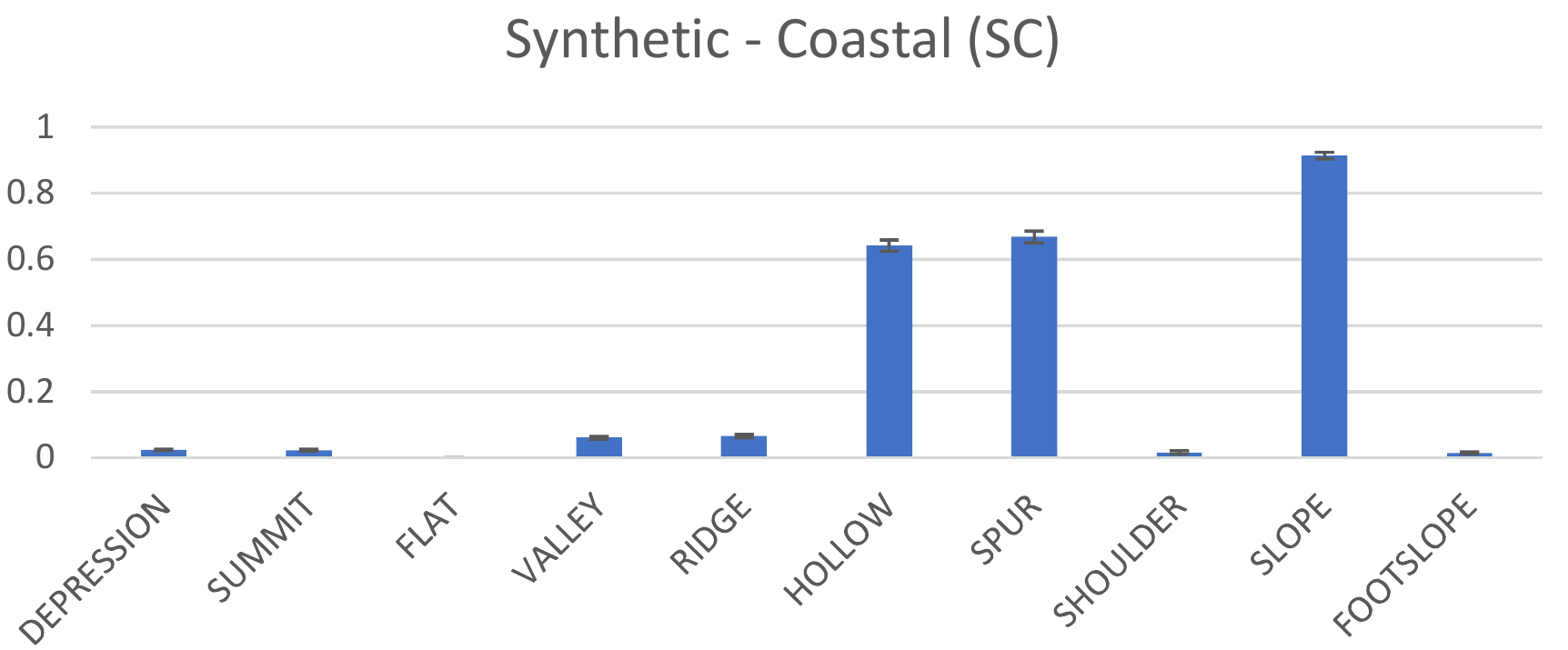}\\
  \includegraphics[width=0.3\linewidth]{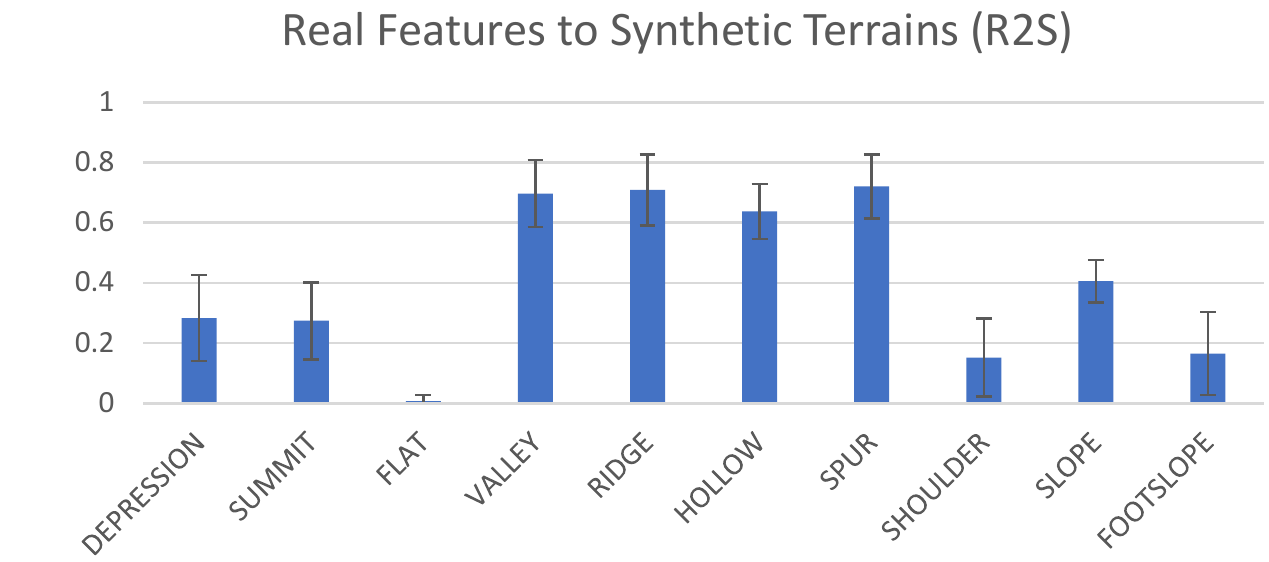}
 \includegraphics[width=0.3\linewidth]{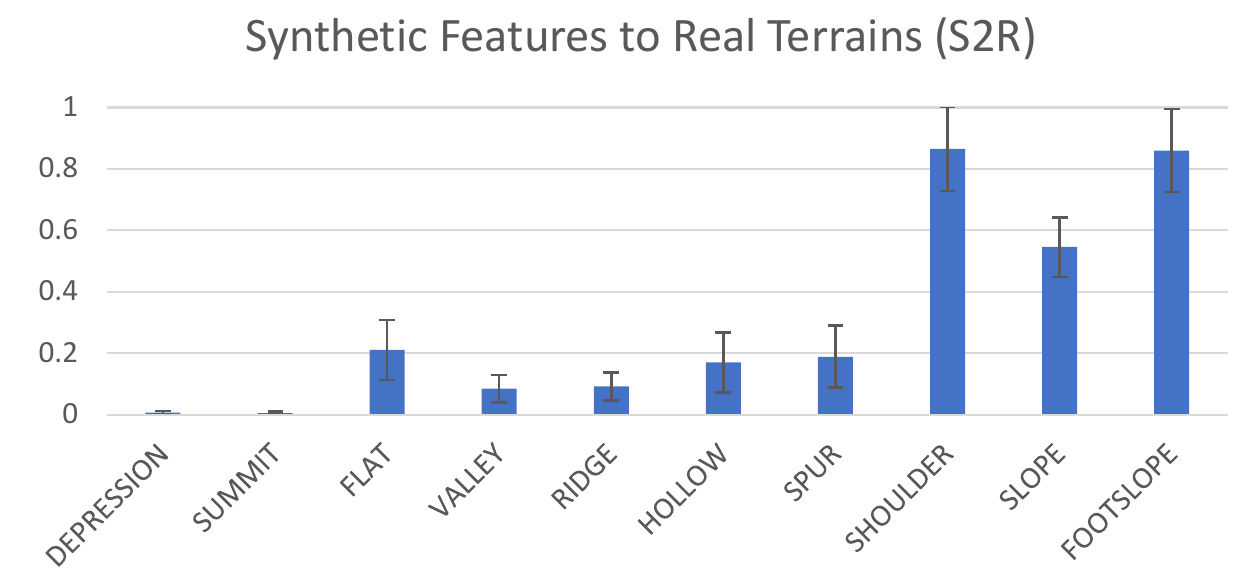}
 \caption{Distribution of the detected geomorphons in real and synthetic terrains from our dataset.}
  \label{fig:geomorphons-graph}
\end{figure*}

Synthetic terrains appear clustered, while features of real terrains are scattered over a wide area. This is further confirmed by the variance of the features as can be seen in graphs in Fig.~\ref{fig:geomorphons-graph}. When the real features are transferred to synthetic terrains, they tend to scatter the images apart and when synthetic features are transferred to real terrains they tend to get close to each other. This seems to indicate that a high variability in geomorphological features is beneficial for perceived realism.

\begin{figure*}[hbt]
  \centering
  \includegraphics[width=\linewidth]{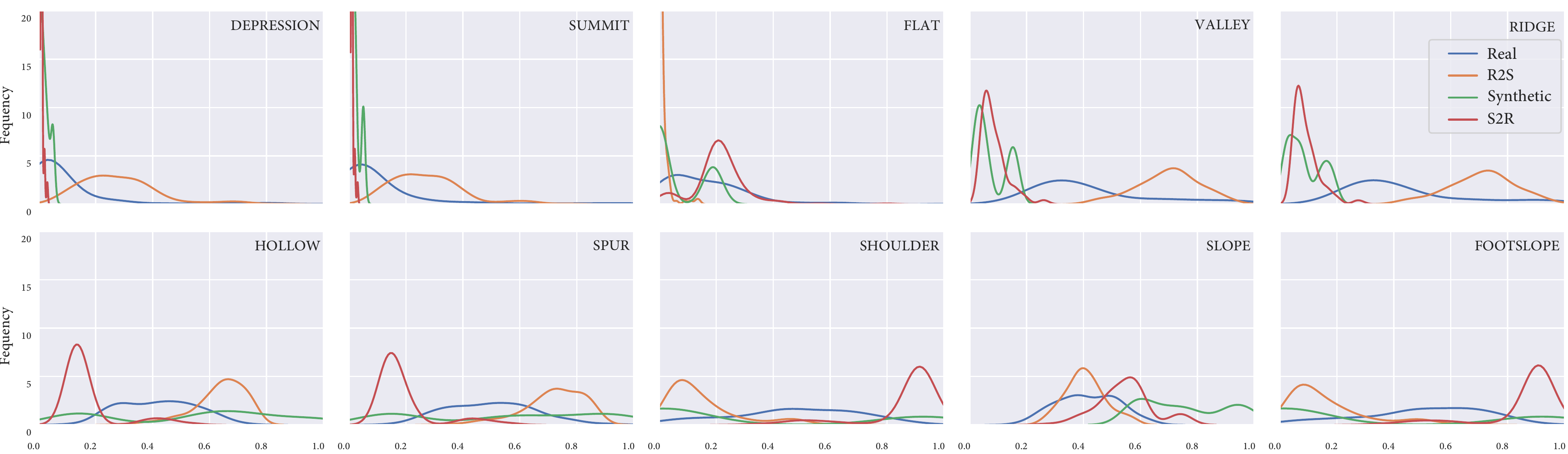}
  \caption{The geomorphon feature comparisons among Real, R2S, Synthetic, and S2R ($x$-axis is the normalized value, the $y$-axis the count).}
  \label{fig:geomorphon_dist}
\end{figure*}

Moreover, we visualize domain-wise comparisons among R, R2S, S, and S2R on the distributions of the element-wise geomorphon feature of terrains in Fig.~\ref{fig:geomorphon_dist}. The geomorphon features of real terrains (blue curve) tend to distribute normally with a wide span. However, the synthetic features (green curve) show significant differences from the real with multi-modal and low-variability distributions on depression, summit, flat, valley, and ridge (Fig.~\ref{fig:geomorphon_dist} top row). We believe the high-peak distributions of synthetic terrains lead to less attractive perceptions than the real. The process of R2S transfer (orange curve) smooths and normalizes the multi-modal high-peak distributions in the synthetic terrains, and improves the perception (refer to Sec.~\ref{sec:perceptual_exp}). It seems that the lack of geomorphon diversity or variability of individual geomorphon feature in distribution may decrease the perceived terrain realism.

\subsection{Perceived Terrain Realism Metrics (PTRM)}\label{sec:PTRM}
The results suggest that the presence of geomorphons is a good indicator of the perceived realism. We devised a Perceived Terrain Realism Metrics (PTRM) that takes as the input a set of normalized geomorphons for a DEM terrain with the spatial resolution of 200m per pixel and returns the estimated perceived realism.

The Pearson correlation coefficients (Tab.~\ref{tab:correlation}) show that there is a strong correlation between each of the geomorphons (our predictor variables) at various levels (Positive and Negative Correlation) on the Perception Score. The order of the influence on the perception score is given by: Valley (0.66), Ridge (0.64), Summit (0.44), Depression (0.42), Spur (0.33), Hollow (0.22), Flat (-0.10), Foot (-0.15), Shoulder (-0.17), and Slope (-0.65).
\begin{table*}[hbt]
\begin{center}
\begin{tabular}{|l|c|c|c|c|c|c|c|c|c|c|c|}
\hline
CORR. & \multicolumn{1}{l|}{DEPR.} & \multicolumn{1}{l|}{SUMM.} & \multicolumn{1}{l|}{FLAT} & \multicolumn{1}{l|}{VALL.} & \multicolumn{1}{l|}{RIDG.} & \multicolumn{1}{l|}{HOLL.} & \multicolumn{1}{l|}{SPUR} & \multicolumn{1}{l|}{SHOU.} & \multicolumn{1}{l|}{SLOP.} & \multicolumn{1}{l|}{FOOT.} & \multicolumn{1}{l|}{SCORE} \\ \hline
DEPR. & 1.00                       & $\bullet$                           & $\bullet$                          & $\bullet$                           & $\bullet$                           & $\bullet$                           & $\bullet$                          & $\bullet$                           & $\bullet$                           & $\bullet$                           & $\bullet$                           \\ \hline
SUMM. & 0.99                       & 1.00                       & $\bullet$                          & $\bullet$                           & $\bullet$                           & $\bullet$                           & $\bullet$                          & $\bullet$                           & $\bullet$                           & $\bullet$                           & $\bullet$                           \\ \hline
FLAT  & -0.41                      & -0.41                      & 1.00                      & $\bullet$                           & $\bullet$                           & $\bullet$                           & $\bullet$                          & $\bullet$                           & $\bullet$                           & $\bullet$                           & $\bullet$                           \\ \hline
VALL. & 0.85                       & 0.87                       & -0.42                     & 1.00                       & $\bullet$                           & $\bullet$                           & $\bullet$                          & $\bullet$                           & $\bullet$                           & $\bullet$                           & $\bullet$                           \\ \hline
RIDG. & 0.86                       & 0.87                       & -0.42                     & 1.00                       & 1.00                       & $\bullet$                           & $\bullet$                          & $\bullet$                           & $\bullet$                           & $\bullet$                           & $\bullet$                           \\ \hline
HOLL. & 0.41                       & 0.42                       & -0.77                     & 0.49                       & 0.49                       & 1.00                       & $\bullet$                          & $\bullet$                           & $\bullet$                           & $\bullet$                           & $\bullet$                           \\ \hline
SPUR  & 0.45                       & 0.46                       & -0.76                     & 0.56                       & 0.57                       & 0.99                       & 1.00                      & $\bullet$                           & $\bullet$                           & $\bullet$                           & $\bullet$                           \\ \hline
SHOU. & -0.50                      & -0.51                      & 0.71                      & -0.53                      & -0.54                      & -0.94                      & -0.93                     & 1.00                       & $\bullet$                           & $\bullet$                           & $\bullet$                           \\ \hline
SLOP. & -0.51                      & -0.53                      & -0.32                     & -0.67                      & -0.66                      & 0.18                       & 0.08                      & -0.18                      & 1.00                       & $\bullet$                           & $\bullet$                           \\ \hline
FOOT  & -0.50                      & -0.50                      & 0.72                      & -0.51                      & -0.52                      & -0.95                      & -0.93                     & 1.00                       & -0.20                      & 1.00                       & $\bullet$                           \\ \hline
SCORE & 0.42                       & 0.44                       & -0.10                     & 0.66                       & 0.64                       & 0.22                       & 0.33                      & -0.17                      & -0.65                      & -0.15                      & 1.00                       \\ \hline
\end{tabular}
\end{center}
\caption{The correlations among ten geomorphons (Depression, Summit, Flat, Valley, Ridge, Hollow, Spur, Shoulder, Slope, and Footslope) and the perception score.}
\label{tab:correlation}
\end{table*}

We performed a multiple linear regression (MLR) model on our dataset with the hypothesis, H\textsubscript{0}: \emph{``There is no linear relationship between the 10 geomorphon landform categories and the perception scores for our terrain data groups.''}
The regression gave us the following statistics: $DFn=10, DFd=588, F = 153.5276, p<0.01$, and with $\alpha=0.01$. Therefore, we rejected the null hypothesis concluding that the coefficients are statistically significant with a $p<0.01$.

\begin{figure}[hbt]
  \centering
  \includegraphics[width=\linewidth]{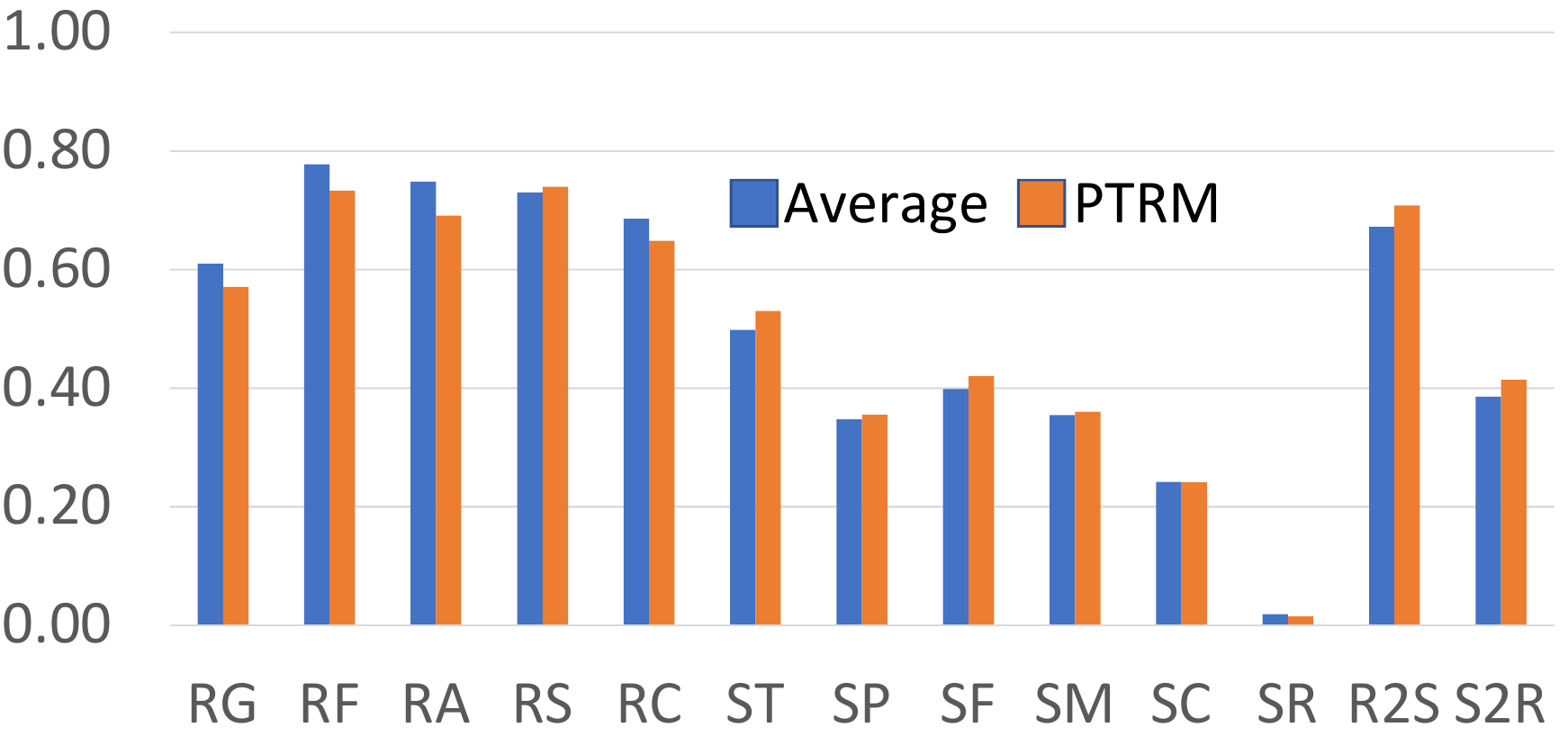}
  \caption{Average value of measured perception score for each category vs. the PTRM.}
  \label{fig:groundtruthgraph}
\end{figure}

The coefficients from the linear regression model between the 10 geomorphons categories are then used to weight the effect of each geomorphon giving the PTRM. The scale for the metrics is $\left<0.0,1.0\right>$ (0=poor, 1=realistic):
\begin{eqnarray}
    PTRM &= &(-38.02 + 3.55G_{depression} + 1.75 G_{summit} +\nonumber \\
    & & 25.12G_{flat}+ 9.61G_{valley} + 7.59G_{ridge} + \\
    & & 6.71G_{hollow} + 9.02G_{spur} + 7.31G_{shoulder} + \nonumber \\
    & & 28.95G_{slope} + 7.63G_{footslope})/69.96.\nonumber
    \label{eqn:PTRM}
\end{eqnarray}
Tab.~\ref{tab:terrain-prediction} and Fig.~\ref{fig:groundtruthgraph} show the comparison of PTRM with the calculated perception score averages for each category.

The resulting R-Squared value for the PTRM is $0.72$ signifying that the~$72\%$ of variation in the visual realism of terrains (\ie\ the perception score) can be explained by the full model with all of our predictor variables \ie~$10$ geomorphon distribution values with a standard error of~$0.13$. All of the landform factors are significant predictors of the perception score.

\begin{table}[hbt]
\begin{center}
\begin{tabular}{ |l | l | c| c |}
\hline
\textbf{Type} & \textbf{Category} & \textbf{Measured Perception Score} & \textbf{PTRM}\\ \hline
Real        & RG  & 0.61  & 0.57\\
(R)         & RF  & 0.78  & 0.73\\
            & RA  & 0.75  & 0.69\\
            & RS  & 0.73  & 0.74\\
            & RC  & 0.69  & 0.65\\ \hline
Synthetic   & ST  & 0.50  & 0.53\\
(S)         & SP  & 0.35  & 0.36\\
            & SF  & 0.40  & 0.42\\
            & SM  & 0.35  & 0.36\\
            & SC  & 0.24  & 0.24\\
            & SR  & 0.02  & 0.02\\  \hline
Transfers   & R2S & 0.67  & 0.71\\
(T)         & S2R & 0.38  & 0.41\\ \hline
\end{tabular}
\end{center}
\caption{A comparison of perception scores generated based on our introduced metrics and our previously normalized score from the study.}\label{tab:terrain-prediction}
\end{table}

\textbf{PTRM Validation:}\label{sec:evaluation}
We have collected a large dataset of various real and generated many synthetic DEMs (Sections~\ref{sec:terrain_data_real} and~\ref{sec:terrain_data_real}). We validated the PTRM by splitting the data five times randomly into 80:20\%, recalculating the PTRM (Eqn~\ref{eqn:PTRM}) on the 80\% and validating on the remaining 20\%. The average regression PTRM for the five dataset is:
\begin{eqnarray}
    PTRM &= &(-38.44 + 3.61 G_{depression} + 1.77 G_{summit} +\nonumber \\
    & & 25.40 G_{flat}+ 9.71 G_{valley} + 7.65G_{ridge} + \\
    & & 6.77G_{hollow} + 9.14 G_{spur} + 7.40 G_{shoulder} + \nonumber \\
    & & 29.26 G_{slope} + 7.69 G_{footslope})/69.22.\nonumber
    \label{eqn:AVGPTRM}
\end{eqnarray}
that is very close to the PTRM model with the amount of explained variation (72\%) and standard error (0.13). Both values remained consistent as the regression model with 95\% confidence interval.

Please note that we also show the PTRM for examples shown in this paper in Figs~\ref{fig:teaser},\ref{fig:synth-real}, \ref{fig:transfer} and all PTRM values for all images as well as perceived scores are in the sumplemental material.

\section{Conclusion}\label{sec:concl}
This paper presented a first step in the direction of evaluating the perceptual quality of procedural models of terrains. We have conducted two large scale perceptual studies that allowed us to rank synthetic and real terrains. Our results show that synthetic terrains are perceived worse than real terrains with statistical significance. We have performed a quantitative study by using geomorphons that indicate that features such as valleys, ridges, summits, depressions, spur, and hollows have significant perceptual importance. We used deep neural network to transfer the features and the second perceptual study confirmed this observation. Eventually, we have designed PTRM that is a novel perceptual metrics based on geomorphons that allows to assign a number of estimated visual quality of the generated terrain.

Our study has several \textbf{limitations}. Geomorphons are localized to small areas of the terrain and they do not reflect the distributions of the large features such as rivers, large valleys, etc. It is possible that two terrains with the same feature vector may be perceived as different because of the variety of distributions. Also, our study made several assumptions on terrain size. Changing the scale of the terrains may have an effect on our results because features of different scales would be captured. Another limitation is the assumption about the terrain classification. While we motivated our classification into terrains with different geomorphological patterns, it is well known that probably every terrain on Earth has been exposed to various morphing phenomena and it is not entirely clear what caused the patterns.
Also, we assumed fixed position of the camera, consistent texturing, and illumination. While these aspects were carefully selected and made constant, it would be interesting to see the effect of each of them on the results. Last but not least, deep feature transfer with GAN provides limited control on the content to be or not to be transferred. With the metrics we provided, the transferred results can be further improved in perception with a better control schema of the generative network. We also did not study the spatial correlation between geomorphons.

There are many possible avenues for \textbf{future work}. Perceptual studies have the potential to answer longstanding questions of visual quality of procedural models. Our work is based on the underlying concept of geomorphons that may be difficult to generalize to different domains. A global metrics considering large geomorphological structures could be also combined with our perceptual study to create another metrics. We intentionally used non-experts to evaluate terrains. it would be interesting to use professional geologists to provide perceptual evaluation. Also, we used only structures commonly found on Earth, non-terrestrial data are increasingly available and it would be interesting to include them as well.

\ifCLASSOPTIONcompsoc
\section*{Acknowledgments}
\else
\section*{Acknowledgment}
\fi
The authors would like to thank Terragen and Vue (e-on Software) for providing student license of their software packages. This research was funded in part by National Science Foundation grants \#10001387, \textit{Functional Proceduralization of 3D Geometric Models}, and project HDW ANR-16-CE33-0001.

\ifCLASSOPTIONcaptionsoff
  \newpage
\fi



\bibliographystyle{IEEEtran}
%
\bibliography{main}



%

\def \aBc{-1.9cm}

\begin{IEEEbiography}
[\vspace{-.7cm}{\includegraphics[height=0.9in,clip,keepaspectratio]
{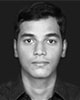}}]{Suren Deepak Rajasekaran}
is a Senior Applied Research Engineer at Sony Corporation of America (SCA). He received his PhD in Technology from the Department of Computer Graphics Technology at Purdue University in 2019. His area of research is in Modeling, Animation and Perception areas of Computer Graphics. 
\end{IEEEbiography}
\vspace{\aBc}

\begin{IEEEbiography}
[\vspace{-.7cm}{\includegraphics[height=0.9in,clip,keepaspectratio]
{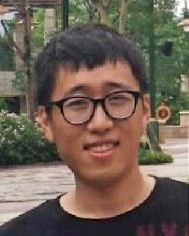}}]{Hao Kang}
is a researcher at Wormpex AI Research and a former member of Purdue High Performance Computer Graphics Laboratory. He received his PhD in Computer Graphics Technology from Purdue University in 2019. His area of research is in 3D computer graphics and vision, and human-robot interaction. 
\end{IEEEbiography}
\vspace{\aBc}

\begin{IEEEbiography}
[\vspace{-.7cm}{\includegraphics[height=0.9in,clip,keepaspectratio]
{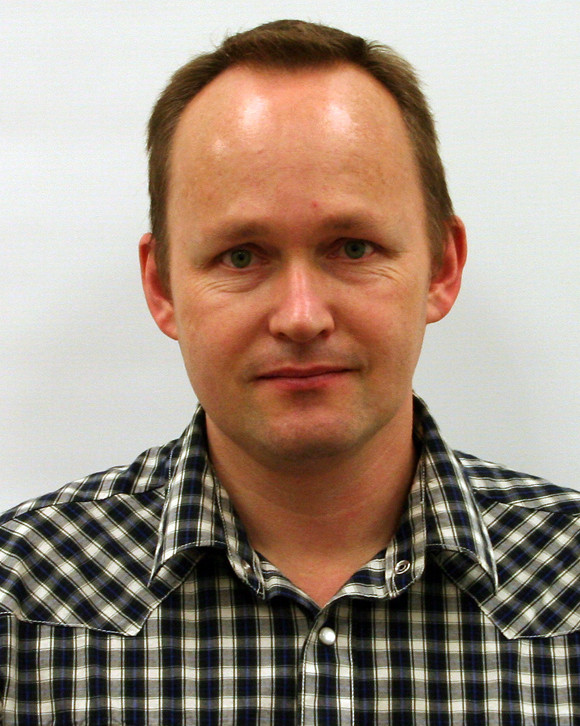}}]{Bedrich Benes}
is George McNelly professor of Technology and professor of Computer Science at Purdue University. His area of research is in procedural and inverse procedural modeling and simulation of natural phenomena and he has published over 150 research papers in the field. 
\end{IEEEbiography}
\vspace{\aBc}

\begin{IEEEbiography}
[\vspace{-.7cm}{\includegraphics[height=0.9in,clip,keepaspectratio]
{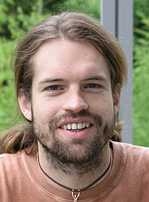}}]{Martin \v{C}ad\'{\i}k}
is an associate professor of computer science at Brno University of Technology, where he heads his Computational Photography group. He received his PhD from Czech Technical University in Prague and his research includes high dynamic range imaging, image processing, computer vision, and image and video quality assessment, among others.
\end{IEEEbiography}
\vspace{\aBc}

\begin{IEEEbiography}
[\vspace{-.7cm}{\includegraphics[height=0.9in,clip,keepaspectratio]
{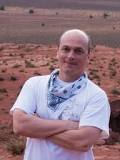}}]{Eric Galin}
is a Professor of Computer Science
at the University Lyon 1 and researcher
at the LIRIS laboratory. He received an engineering degree from Ecole Centrale de Lyon in 1993. His research  in computer graphics include procedural modeling
of virtual worlds, inverse procedural modeling,
simulating natural phenomena and implicit surface modeling.
\end{IEEEbiography}
\vspace{\aBc}

\begin{IEEEbiography}
[\vspace{-.7cm}{\includegraphics[height=0.9in,clip,keepaspectratio]
{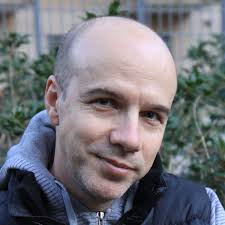}}]{Eric Gu\'erin} is associate professor at INSA Lyon. He received an engineering degree from INSA Lyon in 1998 and a PhD in Computer Science from Universit\'e Lyon 1 in 2002. His research interests include procedural modeling of virtual worlds, natural phenomena and machine learning. He is head of the LIRIS/Geomod team since 2015.
\end{IEEEbiography}
\vspace{\aBc}

\begin{IEEEbiography}
[\vspace{-.7cm}{\includegraphics[height=0.9in,clip,keepaspectratio]
{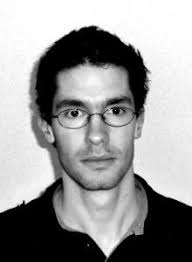}}]{Adrien Peytavie}
is an Assistant  Professor  of Computer  Science  at  the  University  of  Lyon, France. He received a PhD in Computer Science from University Claude Bernard Lyon 1 in Computer Science in 2010. His area of research is in procedural modeling and simulating of virtual worlds.
\end{IEEEbiography}
\vspace{\aBc}

\begin{IEEEbiography}
[\vspace{-.7cm}{\includegraphics[height=0.9in,clip,keepaspectratio]
{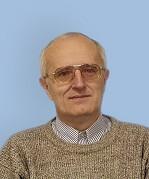}}]{Pavel Slav\'{\i}k}
is a Professor of Computer Science at Czech Technical University in Prague. His area of research is Information Visualization and Human-Computer Interaction. He served as an IPC member in many conferences in these fields. Pavel is one of the founders of the Department of Computer Graphics and Interaction and he co-authored almost 200 publications. 
\end{IEEEbiography}
\vspace{\aBc}





\vfill


\end{document}